\input harvmac
\input epsf
\rightline{EFI-02-72, RI-04-02}   
\Title{
\rightline{hep-th/0204189}
}
{\vbox{\centerline{From Big Bang to Big Crunch and Beyond}}}
\medskip

\centerline{\it S. Elitzur${}^1$, A. Giveon${}^{1}$,
D. Kutasov${}^2$, E. Rabinovici${}^{1}$}
\bigskip
\centerline{${}^1$Racah Institute of Physics, The Hebrew University,
Jerusalem, 91904, Israel}

\centerline{${}^2$Department of Physics, University of Chicago,
5640 S. Ellis Ave., Chicago, IL 60637, USA}

\smallskip

\vglue .3cm
\bigskip

\noindent
We study a quotient Conformal Field Theory, which describes a
$3+1$ dimensional cosmological spacetime. Part of this spacetime
is the Nappi-Witten (NW) universe, which starts at a ``big bang''
singularity, expands and then contracts to a ``big crunch''
singularity at a finite time. The gauged WZW model contains a
number of copies of the NW spacetime, with each copy connected
to the preceeding one and to the next one at the respective big
bang/big crunch singularities. The sequence of NW spacetimes is
further connected at the singularities to a series of non-compact
static regions with closed timelike curves. These regions contain
boundaries, on which the observables of the theory live. This
suggests a holographic interpretation of the physics.

\Date{4/02}

\def\journal#1&#2(#3){\unskip, \sl #1\ \bf #2 \rm(19#3) }
\def\andjournal#1&#2(#3){\sl #1~\bf #2 \rm (19#3) }

\def\ie{{\it i.e.}}
\def\eg{{\it e.g.}}

\noblackbox
%


\def\unlockat{\catcode`\@=11}
\def\lockat{\catcode`\@=12}

\unlockat

\def\newsec#1{\global\advance\secno by1\message{(\the\secno. #1)}
\global\subsecno=0\global\subsubsecno=0\eqnres@t\noindent
{\bf\the\secno. #1}
\writetoca{{\secsym} {#1}}\par\nobreak\medskip\nobreak}
\global\newcount\subsecno \global\subsecno=0
\def\subsec#1{\global\advance\subsecno
by1\message{(\secsym\the\subsecno. #1)}
\ifnum\lastpenalty>9000\else\bigbreak\fi\global\subsubsecno=0
\noindent{\it\secsym\the\subsecno. #1}
\writetoca{\string\quad {\secsym\the\subsecno.} {#1}}
\par\nobreak\medskip\nobreak}
\global\newcount\subsubsecno \global\subsubsecno=0
\def\subsubsec#1{\global\advance\subsubsecno by1
\message{(\secsym\the\subsecno.\the\subsubsecno. #1)}
\ifnum\lastpenalty>9000\else\bigbreak\fi
\noindent\quad{\secsym\the\subsecno.\the\subsubsecno.}{#1}
\writetoca{\string\qquad{\secsym\the\subsecno.\the\subsubsecno.}{#1}}
\par\nobreak\medskip\nobreak}

\def\subsubseclab#1{\DefWarn#1\xdef
#1{\noexpand\hyperref{}{subsubsection}%
{\secsym\the\subsecno.\the\subsubsecno}%
{\secsym\the\subsecno.\the\subsubsecno}}%
\writedef{#1\leftbracket#1}\wrlabeL{#1=#1}}
\lockat

\def\ie{{\it i.e.}}
\def\eg{{\it e.g.}}


\font\manual=manfnt \def\dbend{\lower3.5pt\hbox{\manual\char127}}

\def\IZ{\relax\ifmmode\mathchoice
{\hbox{\cmss Z\kern-.4em Z}}{\hbox{\cmss Z\kern-.4em Z}}
{\lower.9pt\hbox{\cmsss Z\kern-.4em Z}}
{\lower1.2pt\hbox{\cmsss Z\kern-.4em Z}}\else{\cmss Z\kern-.4em
Z}\fi}
\def\half{{1\over 2}}


\def\IZ{\relax\ifmmode\mathchoice
{\hbox{\cmss Z\kern-.4em Z}}{\hbox{\cmss Z\kern-.4em Z}}
{\lower.9pt\hbox{\cmsss Z\kern-.4em Z}}
{\lower1.2pt\hbox{\cmsss Z\kern-.4em Z}}\else{\cmss Z\kern-.4em
Z}\fi}
\def\IB{\relax{\rm I\kern-.18em B}}
\def\IC{{\relax\hbox{$\inbar\kern-.3em{\rm C}$}}}
\def\ID{\relax{\rm I\kern-.18em D}}
\def\IE{\relax{\rm I\kern-.18em E}}
\def\IF{\relax{\rm I\kern-.18em F}}
\def\IG{\relax\hbox{$\inbar\kern-.3em{\rm G}$}}
\def\IGa{\relax\hbox{${\rm I}\kern-.18em\Gamma$}}
\def\IH{\relax{\rm I\kern-.18em H}}
\def\II{\relax{\rm I\kern-.18em I}}
\def\IK{\relax{\rm I\kern-.18em K}}
\def\IP{\relax{\rm I\kern-.18em P}}
\def\IQ{\relax\hbox{$\inbar\kern-.3em{\rm Q}$}}

\def\inbar{\,\vrule height1.5ex width.4pt depth0pt}

\font\cmss=cmss10 \font\cmsss=cmss10 at 7pt
\def\IR{\relax{\rm I\kern-.18em R}}

\def\Tr{\rm Tr}

%
%

\def\makeblankbox#1#2{\hbox{\lower\dp0\vbox{\hidehrule{#1}{#2}%
   \kern -#1
   \hbox to \wd0{\hidevrule{#1}{#2}%
      \raise\ht0\vbox to #1{}
      \lower\dp0\vtop to #1{}
      \hfil\hidevrule{#2}{#1}}%
   \kern-#1\hidehrule{#2}{#1}}}%
}%
\def\hidehrule#1#2{\kern-#1\hrule height#1 depth#2 \kern-#2}%
\def\hidevrule#1#2{\kern-#1{\dimen0=#1\advance\dimen0 by #2\vrule
    width\dimen0}\kern-#2}%
\def\openbox{\ht0=1.2mm \dp0=1.2mm \wd0=2.4mm  \raise 2.75pt
\makeblankbox {.25pt} {.25pt}  }

\def\bun#1/#2{\leavevmode
   \kern.1em \raise .5ex \hbox{\the\scriptfont0 #1}%
   \kern-.1em $/$%
   \kern-.15em \lower .25ex \hbox{\the\scriptfont0 #2}%
}

\def\opensquare{\ht0=3.4mm \dp0=3.4mm \wd0=6.8mm  \raise 2.7pt
\makeblankbox {.25pt} {.25pt}  }


\def\sector#1#2{\ {\scriptstyle #1}\hskip 1mm
\mathop{\opensquare}\limits_{\lower 1mm\hbox{$\scriptstyle#2$}}\hskip 1mm}

\def\tsector#1#2{\ {\scriptstyle #1}\hskip 1mm
\mathop{\opensquare}\limits_{\lower 1mm\hbox{$\scriptstyle#2$}}^\sim\hskip 1mm}


\def\inbar{\,\vrule height1.5ex width.4pt depth0pt}

\font\cmss=cmss10 \font\cmsss=cmss10 at 7pt
\def\IR{\relax{\rm I\kern-.18em R}}

\def\Tr{\rm Tr}


\def\frac#1#2{{#1\over#2}}

\def\half{\frac12}

\def\inbar{\,\vrule height1.5ex width.4pt depth0pt}
\def\IC{\relax\hbox{$\inbar\kern-.3em{\rm C}$}}
\def\IR{\relax{\rm I\kern-.18em R}}
\def\IP{\relax{\rm I\kern-.18em P}}

%
%
\catcode`\@=11
\def\slash#1{\mathord{\mathpalette\c@ncel{#1}}}
\overfullrule=0pt

\def\II{{\cal I}}

\def\MM{{\cal M}}

\def\underrel#1\over#2{\mathrel{\mathop{\kern\z@#1}\limits_{#2}}}

\catcode`\@=12


%

\def\sinh{{\rm sinh}}
\def\cosh{{\rm cosh}}

\def\exp{{\rm exp}}



\def\frac#1#2{{#1\over#2}}

\def\half{\frac12}

\def\inbar{\,\vrule height1.5ex width.4pt depth0pt}
\def\IC{\relax\hbox{$\inbar\kern-.3em{\rm C}$}}
\def\IR{\relax{\rm I\kern-.18em R}}
\def\IP{\relax{\rm I\kern-.18em P}}

%
%

%
\catcode`\@=11
\def\slash#1{\mathord{\mathpalette\c@ncel{#1}}}
\overfullrule=0pt

\def\II{{\cal I}}

\def\MM{{\cal M}}

\def\underrel#1\over#2{\mathrel{\mathop{\kern\z@#1}\limits_{#2}}}

\catcode`\@=12


%

\def \sinh{{\rm sinh}}
\def \cosh{{\rm cosh}}

\def\exp{{\rm exp}}


\lref\BanksYP{
T.~Banks and W.~Fischler,
``M-theory observables for cosmological space-times,''
arXiv:hep-th/0102077.
}

\lref\vilenkin{N.J. Vilenkin, ``Special Functions and the Theory
of Group Representations,'' AMS, 1968.}

\lref\NappiKV{
C.~R.~Nappi and E.~Witten,
``A Closed, expanding universe in string theory,''
Phys.\ Lett.\ B {\bf 293}, 309 (1992)
[arXiv:hep-th/9206078].
}

\lref\GiveonFU{
A.~Giveon, M.~Porrati and E.~Rabinovici,
``Target space duality in string theory,''
Phys.\ Rept.\  {\bf 244}, 77 (1994)
[arXiv:hep-th/9401139].
}

\lref\GiveonPH{
A.~Giveon and E.~Kiritsis,
``Axial vector duality as a gauge symmetry and 
topology change in string theory,''
Nucl.\ Phys.\ B {\bf 411}, 487 (1994)
[arXiv:hep-th/9303016].
}

\lref\GiveonKB{
A.~Giveon and A.~Pasquinucci,
``On cosmological string backgrounds with toroidal isometries,''
Phys.\ Lett.\ B {\bf 294}, 162 (1992)
[arXiv:hep-th/9208076].
}

\lref\KiritsisFD{
E.~Kiritsis and C.~Kounnas,
``Dynamical topology change in string theory,''
Phys.\ Lett.\ B {\bf 331}, 51 (1994)
[arXiv:hep-th/9404092].
}

\lref\HoravaAM{
P.~Horava,
``Some exact solutions of string 
theory in four-dimensions and five-dimensions,''
Phys.\ Lett.\ B {\bf 278}, 101 (1992)
[arXiv:hep-th/9110067].
}

\lref\KounnasWC{
C.~Kounnas and D.~Lust,
``Cosmological string backgrounds from gauged WZW models,''
Phys.\ Lett.\ B {\bf 289}, 56 (1992)
[arXiv:hep-th/9205046].
}

\lref\agmoo{
O.~Aharony, S.~S.~Gubser, J.~Maldacena, H.~Ooguri and Y.~Oz,
``Large N field theories, string theory and gravity,''
Phys.\ Rept.\  {\bf 323}, 183 (2000)
[arXiv:hep-th/9905111].
}

\lref\devega{
H.~J.~de Vega, A.~L.~Larsen and N.~Sanchez,
``Non-singular string-cosmologies from exact conformal field theories,''
Nucl.\ Phys.\ Proc.\ Suppl.\  {\bf 102}, 201 (2001)
[arXiv:hep-th/0110262].
}

\lref\brs{
K.~Bardakci, E.~Rabinovici and B.~Saering,
``String Models With $C<1$ Components,''
Nucl.\ Phys.\ B {\bf 299}, 151 (1988).
}

\lref\hawel{S.W. Hawking and G.F. Ellis, ``The large scale structure
of space-time'' Cambridge University Press 1976}

\lref\AharonyUB{
O.~Aharony, M.~Berkooz, D.~Kutasov and N.~Seiberg,
``Linear dilatons, NS5-branes and holography,''
JHEP {\bf 9810}, 004 (1998)
[arXiv:hep-th/9808149].
}

\lref\KhouryBZ{
J.~Khoury, B.~A.~Ovrut, N.~Seiberg, P.~J.~Steinhardt and N.~Turok,
``From big crunch to big bang,''
arXiv:hep-th/0108187.
}

\lref\CornalbaFI{
L.~Cornalba and M.~S.~Costa,
``A New Cosmological Scenario in String Theory,''
arXiv:hep-th/0203031.
}

\lref\NekrasovKF{
N.~A.~Nekrasov,
``Milne universe, tachyons, and quantum group,''
arXiv:hep-th/0203112.
}

\lref\SimonMA{
J.~Simon,
``The geometry of null rotation identifications,''
arXiv:hep-th/0203201.
}

\lref\lms{H.~Liu, G.~Moore and N.~Seiberg,
``Strings in a Time-Dependent Orbifold,''
 arXiv:hep-th/0204168.
 }

\lref\BarsDX{
I.~Bars and K.~Sfetsos,
``$SL(2,\IR) \times SU(2) / \IR^2$ 
string model in curved space-time and exact 
conformal results,''
Phys.\ Lett.\ B {\bf 301}, 183 (1993)
[arXiv:hep-th/9208001].
}

\newsec{Introduction}

Defining observables in quantum gravity
in cosmological spacetimes is an interesting
problem that received some attention over the years. In particular,
in the presence of initial (big bang) and/or final (big crunch)
singularities, one might expect that the initial and/or final 
conditions which give rise to observables
must be specified at or near a singularity. This is puzzling 
since in more familiar situations, such as asymptotically
flat spacetimes (with or without a spacelike linear dilaton),
and asymptotically AdS spacetimes, boundary conditions (observables)
are specified in regions where the theory becomes simple in appropriate
variables, whereas near a cosmological singularity the interactions
are expected to be strong. It might be that quantum gravity in fact
simplifies near such singularities \BanksYP, but a satisfactory
demonstration of that is currently lacking.

\lref\WittenKN{
E.~Witten,
``Quantum gravity in de Sitter space,''
arXiv:hep-th/0106109.
}

\lref\bcr{
K.~Bardacki, M.~J.~Crescimanno and E.~Rabinovici,
``Parafermions From Coset Models,''
Nucl.\ Phys.\ B {\bf 344}, 344 (1990).
}

\lref\BrusteinKW{
R.~Brustein and G.~Veneziano,
``The Graceful exit problem in string cosmology,''
Phys.\ Lett.\ B {\bf 329}, 429 (1994)
[arXiv:hep-th/9403060].
}

\lref\VenezianoPZ{
G.~Veneziano,
``String cosmology: The pre-big bang scenario,''
arXiv:hep-th/0002094.
}

It is natural to enquire whether string theory sheds
new light on these issues. The purpose of this paper is to 
study this question in a particular model introduced by
Nappi and Witten \NappiKV. Related four dimensional 
backgrounds appeared earlier in \refs{\HoravaAM,\KounnasWC}. 
They all belong to a moduli space of cosmological string
backgrounds with two Abelian isometries \GiveonKB.

The original Nappi-Witten (NW) model describes a four
dimensional anisotropic
closed universe, which starts from a big bang, and recollapses
to a big crunch after a finite time. Therefore, 
the problem of defining observables is particularly acute
in this case \WittenKN. At the same time, the authors of
\NappiKV\ found that this cosmology is described by a certain
coset CFT, $[SL(2)\times SU(2)]/U(1)^2$, which can be studied
using the tools of weakly coupled string theory. In particular, 
one expects to be able to define observables in the standard 
fashion, using vertex operators for describing perturbative
string states. 

\lref\GasperiniEM{
M.~Gasperini and G.~Veneziano,
``Pre - big bang in string cosmology,''
Astropart.\ Phys.\  {\bf 1}, 317 (1993)
[arXiv:hep-th/9211021].
}

\lref\TolleyCV{
A.~J.~Tolley and N.~Turok,
``Quantum fields in a big crunch / big bang spacetime,''
arXiv:hep-th/0204091.
}

It is thus interesting to ask what is the interpretation of
the observables one finds in string theory on the
NW background in terms of the big bang/big crunch cosmology.
String theory resolves this problem in an interesting way.
The coset model that gives rise to the NW cosmology describes
a spacetime which contains additional regions, which we will
refer to as {\it whiskers} (see figure 3 below).
These regions are connected to the cosmological spacetime
discussed in \NappiKV\ at the big bang and big crunch
singularities. Moreover, the coset contains a number of copies
of the NW spacetime, attached to each other at the respective
big bang/big crunch singularities. As we will see below,
the different regions are coupled by the
dynamics. Thus, this model falls into the pre-big bang class.
Other models in this class were discussed \eg\ in
\refs{\GasperiniEM,\BrusteinKW} (for a review, see \VenezianoPZ),
and more recently in
\refs{\KhouryBZ,\CornalbaFI,\NekrasovKF,\SimonMA,\TolleyCV,\lms}.

The whiskers are non-compact, time independent and contain a boundary
at spatial infinity, near which the background asymptotes to a
spacelike linear dilaton one. As in \AharonyUB, it is natural to
define the observables of the model via the behavior of fields near
this boundary. The vertex operators describing perturbative strings
living on the coset correspond to such observables.

Correlation functions of these observables provide information
about physics in the bulk of spacetime. We discuss two classes
of observables: those that describe scattering states, which correspond
to wavefunctions that are $\delta$-function normalizable near the
boundary, and non-normalizable wavefunctions localized at the
boundary, that are similar to the standard observables in AdS.

Correlation functions of the $\delta$-function normalizable
observables provide information about scattering of states that
live near the boundary from the singularities.
One finds that incoming waves may be partially reflected back
into the non-compact region, and partially transmitted into the
NW cosmological region. They might then reemerge in another
asymptotic region. Correlation functions of non-normalizable
observables provide information about states living in the NW
cosmology.

An interesting aspect of the coset CFT is that the whiskers
contain closed timelike curves. At first sight this seems
to be a major obstacle for including them in the geometry.
Nevertheless, they seem to be needed for defining observables,
and as we will see later, the dynamics couples the NW cosmological
region to the whiskers. It is thus natural to wonder whether
the model is physically consistent. This is expected to be the
case on general grounds, since the model is obtained by a seemingly
sensible gauging of a consistent string background. An observation 
which might
be relevant for this issue is that, technically, the reason for the
occurrence of closed timelike curves in the model is the compactness
of the $SU(2)$ component of the coset. This compactness is also
responsible for the compactness of the NW cosmological region,
and for a depletion in the spectrum of perturbative string states.
Thus, one may hope that the geometry, spectrum and interactions are 
precisely such that the model is physically consistent (\eg\ free from
violations of causality). One of the motivations for this study
is to find out whether this is indeed the case.

Another  interesting aspect of the physics of the NW model is that the
presence of the big bang and big crunch singularities does not
appear to lead to a breakdown of string perturbation theory. The
correlation functions of vertex operators appear to be well behaved,
at least at low orders of string perturbation theory. One possible
qualitative interpretation of this is that stringy probes are smeared
over distances of order the string scale, and thus are able to pass
through the big bang and big crunch singularities. Also, it turns out
that the NW spacetime is non-Hausdorff near the singularity. This too
may play a role in resolving the apparent singularities in the geometry.
To shed more light on the behavior near the singularity, it would be
interesting to study the dynamics of D-branes in this geometry.

In the remainder of the paper we describe in more detail
the structure of the NW model, and make some of the above 
assertions more precise. The plan of the paper is as follows.
In section 2 we discuss the $SL(2,\IR)$ group manifold. 
For application to the NW model, it is necessary to 
diagonalize a spacelike non-compact one dimensional 
subgroup of $SL(2,\IR)$. We discuss the geometry and 
the behavior of eigenfunctions of the above $U(1)$ subgroup.
This is a well studied mathematical problem \vilenkin. 
One finds that the $SL(2,\IR)$ group manifold splits into 
different regions. Eigenfunctions of the non-compact spacelike 
$U(1)$ are generically non-analytic at the boundaries between 
different regions.

In section 3 we turn to the NW coset. We describe the modifications
of the group topology due to the identifications, and
the geometrical data corresponding to the coset in different regions.
Some regions describe the closed NW cosmology, while others 
are static and contain asymptotic regimes, closed timelike curves and 
timelike singularities. We determine the vertex operators of low lying 
string states corresponding to the principal continuous and discrete 
series representations of $SL(2, \IR)$. We present examples of scattering 
amplitudes which demonstrate that fundamental
strings may go from one region to another in the NW background.
We also show that the compactness of $SU(2)$ implies a large
depletion in the spectrum of physical states in the model.

In section 4 we discuss the quantization of the superstring
in the NW background. The theory appears to be non-supersymmetric,
but one nevertheless needs to perform a chiral GSO projection.
We describe the projection and some aspects of the spectrum
of the theory. Our main results are summarized in section 5.

\newsec{$SL(2,\IR)$}

\subsec{Geometry}

The NW model \NappiKV\ inherits much of its structure
from the underlying CFT on $SL(2,\IR)$. Therefore, in this
section we describe some properties of the $SL(2,\IR)$ group
manifold in variables that will be useful when we study
the gauged WZW model.

One of the characteristic geometrical features of the coset model
will turn out to be its decomposition into two different
types of regions. There are regions compact in space and time
which can be interpreted as a succession of expanding and
contracting universes. There are also non-compact, static regions
extending all the way to infinity. This decomposition can
be traced back to the geometry of $SL(2,\IR)$.

A general group element $g$ in $SL(2,\IR)$ has the form
\eqn\defho{g=\left(\matrix{
a&b\cr
c&d\cr}\right)~,}
where
\eqn\abcd{ad-bc=1; \qquad a,b,c,d\in \IR~.}
The $SL(2,\IR)_L\times SL(2,\IR)_R$ symmetry of the
WZW model on the group manifold acts on $g$ via left
and right multiplication by independent $SL(2,\IR)$
matrices. The NW model is obtained by gauging a
group containing the
spacelike non-compact $U(1)_L\times U(1)_R$ symmetry
generated by $\sigma_3$. This acts on $g$ as
\eqn\uoneact{g\to e^{\alpha\sigma_3} g e^{\beta\sigma_3}
=
\left(\matrix{
ae^{\alpha+\beta}&be^{\alpha-\beta}\cr
ce^{\beta-\alpha}&de^{-\alpha-\beta}\cr}\right)~.}
In a unitary representation,
eigenfunctions of $U(1)_L\times U(1)_R$ transform under
the action \uoneact\ as
\eqn\eignef{D_{m,\bar m}\to e^{2(im\alpha+i\bar m\beta)} D_{m,\bar m}~,}
with real $m, \bar m$.
Viewed as functions on the group manifold, they have the form
\eqn\dmbarm{D_{m,\bar m}=
\left({a\over d}\right)^{{i\over2}(m+\bar m)}
\left({b\over c}\right)^{{i\over2}(m-\bar m)}K(ad)~,}
where $K$ is a function to be determined.
Therefore, we see that eigenfunctions of $U(1)_L\times U(1)_R$
typically exhibit non-analytic behavior when one of the elements of the
matrix \defho\ goes to zero. Thus, the $SL(2,\IR)$ group manifold
naturally splits into different regions, depending on the signs
of $a,b,c,d$.
There are twelve different regions:
\item{(A)} $ad>0$, $bc>0$. There are four such regions,
depending on the two signs, sign$(a)=$ sign$(d)$ and
sign$(b)=$ sign$(c)$.
\item{(B)} $ad>0$, $bc<0$ (four regions).
\item{(C)} $ad<0$, $bc<0$ (four regions).

\noindent
The different regions can be distinguished by the value of the
quantity
\eqn\defw{W={\rm Tr} (\sigma_3 g\sigma_3 g^{-1})=2(2ad-1)=2(2bc+1)~,}
which is invariant under the $U(1)_L\times U(1)_R$ symmetry
\uoneact. In regions (A), $W>2$; in regions (B), $2>W>-2$;
in regions (C), $W<-2$.

\bigskip
{\vbox{{\epsfxsize=3in
        \nobreak
    \centerline{\epsfbox{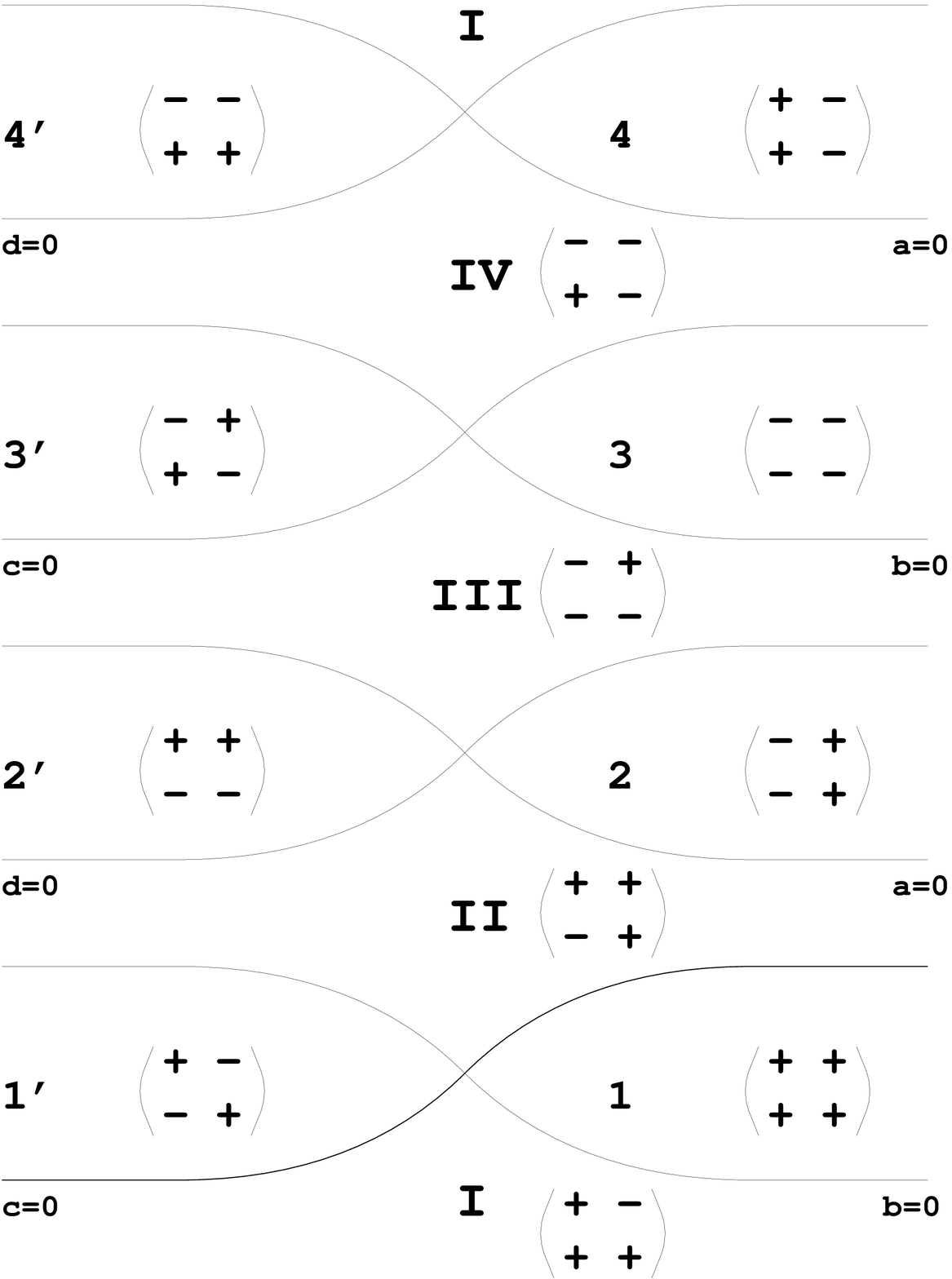}}
        \nobreak\bigskip
    {\raggedright\it \vbox{
{\bf Figure 1.}
{\it
The $SL(2,\IR)$ group manifold split into regions discussed in the text.
} }}}}
    \bigskip}
\noindent
In figure 1 we illustrate the resulting structure. The signs
indicated in the figure are those of
$\left(\matrix{a&b\cr c&d}\right)$. The regions
$1$, $1'$, $3$, $3'$ are of type (A) in the classification above,
while $2$, $2'$, $4$, $4'$ are of type (C). Regions $I$, $II$,
$III$, $IV$ are the four regions of type (B). In string theory
on $AdS_3$ (the universal cover of $SL(2,\IR)$), 
a special role is played by the boundary of the
$SL(2,\IR)$ group manifold, which is the space on which the
dual CFT is defined. The boundary corresponds to large $a,b,c,d$,
or $|W|\to\infty$ (see \defw). In figure 1 this corresponds to
the asymptotic infinities in regions $1-4$, $1'-4'$. The regions
$I$, $II$, $III$, $IV$ do not reach the boundary of $AdS_3$.

After gauging \NappiKV, regions of type (B), where $|W|\le 2$,
give rise to NW cosmologies. The others give rise to non-compact, static
regions with closed timelike curves.


One can restrict attention to $PSL(2,\IR)$, which is obtained
by identifying $g$ with $-g$ in \defho\ (a symmetry of \abcd).
Using this symmetry to (say) set $a>0$, one is left with the
six regions $1$, $1'$, $2'$, $4$, $I$, $II$, as indicated in figure 2.

\bigskip
{\vbox{{\epsfxsize=3in
        \nobreak
    \centerline{\epsfbox{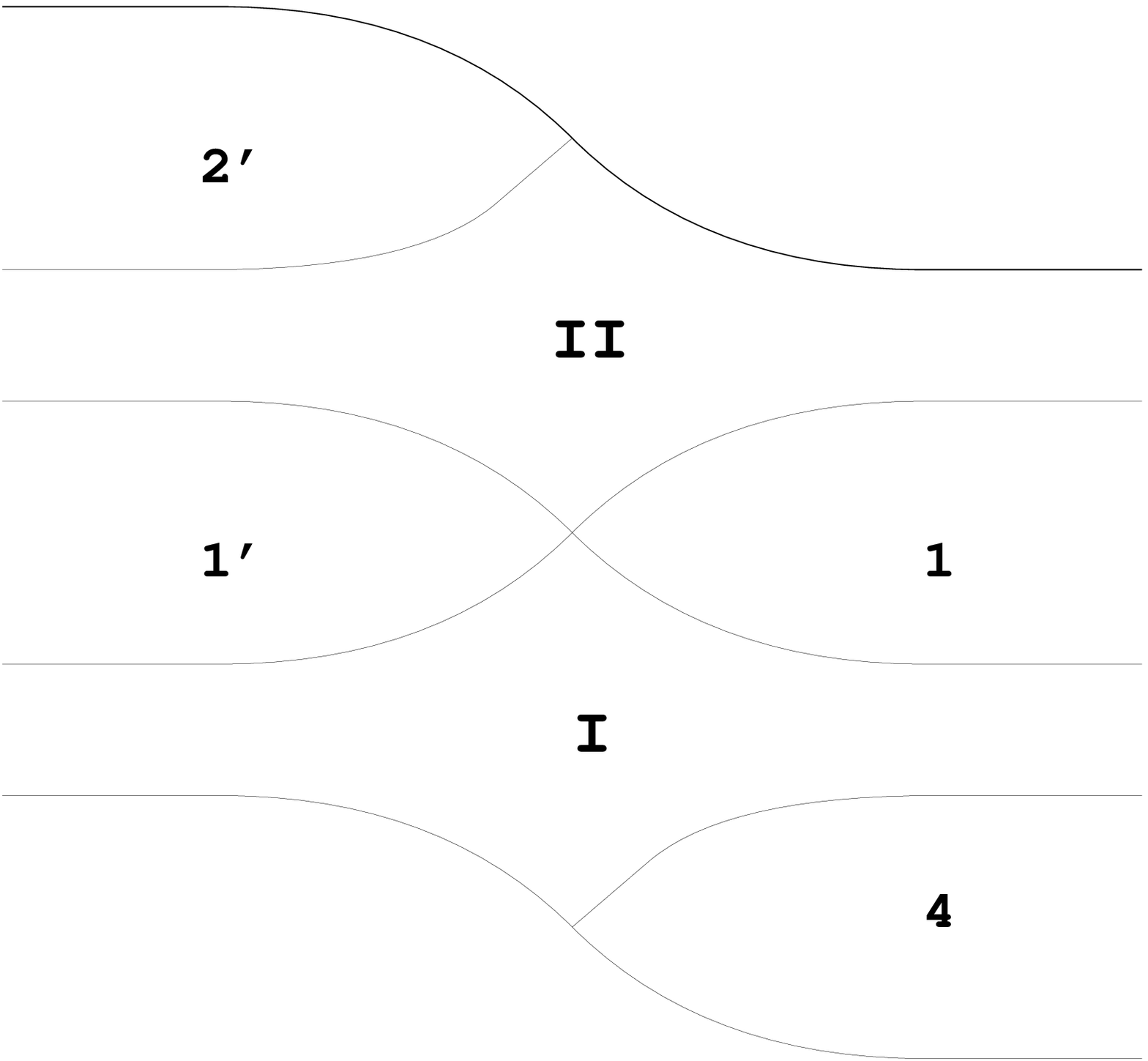}}
        \nobreak\bigskip
    {\raggedright\it \vbox{
{\bf Figure 2.}
{\it
The geometry of $PSL(2,\IR)$, or the Poincare patch.
} }}}}
    \bigskip}
\noindent
The asymptotic region in $PSL(2,\IR)$ (the boundary) is connected,
and forms a conformal compactification of two dimensional Minkowski
spacetime.

The $PSL(2,\IR)$ group manifold as well as its double cover
$SL(2,\IR)$ are not simply connected; they have
closed timelike curves. This is reflected in the periodic
identification in the vertical direction in figures 1,2.
This identification is usually avoided
by considering the universal cover of the group manifold.
An element of the universal cover corresponding to a $g\in SL(2,\IR)$
is (the homotopy class of) a curve  starting at the
identity and ending at $g$. Since $\pi_1(SL(2,\IR))=\IZ$, there is an
infinite number of elements in the universal cover corresponding to
every $g\in SL(2,\IR)$. In the universal cover, the region shown
in figure 2 is known as the Poincare patch.

The Poincare patch plays an important role for a number of reasons.
First, it is naturally obtained by analytic continuation from
Euclidean $AdS_3$. Indeed, the Euclidean version of \abcd\ is
the hyperboloid
\eqn\euclid{X_0^2=X_1^2+X_2^2+X_3^2+1~.}
The space \euclid\ has two disconnected components, corresponding
to positive and negative $X_0$. Euclidean $AdS_3$ corresponds to
one component. It can be parametrized as follows:
\eqn\ugg{\left(\matrix{
a&b\cr
c&d\cr}\right)=
\left(\matrix{
X_0+X_1&X_2+iX_3\cr
X_2-iX_3&X_0-X_1\cr}\right)=
\left(\matrix{
u&u\gamma^+\cr
u\gamma^-&{1\over u}+u\gamma^+\gamma^-\cr}\right)~,
}
where the Poincare coordinates $u, \gamma^\pm$ satisfy
$u\in \IR_+$, $\gamma^+\in \IC$, and $\gamma^-=
(\gamma^+)^*$. The analytic continuation to the
Lorentzian manifold is obtained by taking $\gamma^+$
and $\gamma^-$ to be independent real (lightlike) variables.
The hyperboloid $u\ge0$ maps to the region $a\ge0$ in $SL(2,\IR)$
(see \ugg). Since there are no identifications on the
coordinates, the resulting Lorentzian manifold should be thought
of as a patch in the universal cover, and not as $PSL(2,\IR)$.
This space, the Poincare patch, corresponds to figure 2 without the
vertical identifications.

The $SL(2)$ invariant line element on the Poincare patch is
\eqn\lineel{ds^2={du^2\over u^2}+u^2d\gamma^+d\gamma^-~.}
Another reason to consider the Poincare patch is that the metric
\lineel\ is obtained in the near-horizon geometry
of branes in string theory. For example, systems of fundamental strings
and $NS5$-branes naturally give this geometry
(for a review, see \agmoo).

\subsec{Wavefunctions}

\lref\tesc{
J.~Teschner,
``On structure constants and fusion rules in the SL(2,C)/SU(2) WZNW  model,''
Nucl.\ Phys.\ B {\bf 546}, 390 (1999)
[arXiv:hep-th/9712256].
}

The vertex operators on the coset are obtained by a restriction of
those on the group manifold.
Therefore, we will describe in this subsection the explicit forms of
eigenfunctions of the Laplacian on $SL(2,\IR)$ in the basis
\eignef, \dmbarm.
We are mainly interested in the Minkowski problem, but since
on a Euclidean worldsheet the Euclidean $AdS_3(=H_3^+)$ CFT
is better behaved, we first describe the Euclidean analogs
of these wavefunctions \tesc.

One starts with the well known eigenfunctions of the Laplacian
on $AdS_3$,
\eqn\fheig{\Phi_h(x,\bar x;\phi,\gamma,\bar\gamma)=
\left(|\gamma-x|^2e^\phi+e^{-\phi}\right)^{-2h}~,}
where the coordinates $\phi,\gamma,\bar\gamma$ are related
to those in \ugg, \lineel\ by $u=e^\phi$; $\gamma=\gamma^+$;
$\bar\gamma=\gamma^-$. The corresponding eigenvalue of the Laplacian
is $-h(h-1)$.

\lref\GiveonNS{
A.~Giveon, D.~Kutasov and N.~Seiberg,
``Comments on string theory on AdS(3),''
Adv.\ Theor.\ Math.\ Phys.\  {\bf 2}, 733 (1998)
[arXiv:hep-th/9806194];
}

\lref\KutasovXU{
D.~Kutasov and N.~Seiberg,
``More comments on string theory on AdS(3),''
JHEP {\bf 9904}, 008 (1999)
[arXiv:hep-th/9903219].
}

The generators of $SL(2)$ are realized on $\Phi_h$ as the
differential operators
\eqn\jdiffops{\eqalign{
J^3=&\gamma\partial_\gamma-\half\partial_\phi=-(x\partial_x+h)~,\cr
J^-=&\partial_\gamma=-\partial_x~,\cr
J^+=&\gamma^2\partial_\gamma-\gamma\partial_\phi-e^{-2\phi}
\partial_{\bar\gamma}=-(x^2\partial_x+2hx)~.\cr
}}
Similar formulae hold for the right moving generators $\bar J^a$.
$\bar J^3$ generates the transformation $g\to g e^{\beta\sigma_3}$
in \uoneact.  This symmetry acts as $u\to e^\beta u$,
$\gamma\to e^{-2\beta}\gamma$, which corresponds to
(holomorphic) dilation symmetry on the boundary of $AdS_3$.
Thus, $J^3$ can be thought of as $L_0$ of the Virasoro algebra
acting on the boundary of $AdS$ space 
\refs{\GiveonNS,\KutasovXU}. In that 
interpretation, one considers wavefunctions which have $L_0\in \IR$.
For our purposes, we saw in eq. \eignef\ that one needs
to consider wavefunctions with $J^3\in i\IR$. This has to do
with the fact that in the analytic continuation from Euclidean
to Minkowski spacetime corresponding to $X_3\to iX_3$ in \ugg,
one finds that the timelike generator of $SL(2,\IR)$ is in fact
$-iJ^2=J^--J^+$, and $-iJ^3$ becomes a spacelike generator, with
real eigenvalues.

Eigenstates of $J^3$ with imaginary eigenvalues have the (formal) form
\eqn\khh{K_{m,\bar m;j}=\int d^2 x x^{j+im}\bar x^{j+i\bar m}
\Phi_{j+1}(x,\bar x;\phi,\gamma,\bar\gamma)~.}
They are the same as the operators that are usually considered
in CFT on $AdS_3$, with $m$ replaced by $im$.

We now return to the Minkowski problem, which is richer, since
one has to study the wavefunctions in the different regions\foot{In
Euclidean space, comparing \dmbarm\ and \ugg, one finds that
$a/d$, $b/c$ never vanish on $H_3^+$, and thus there is no analog
of the different regions appearing in the Minkowski case.} in figures 1,2.
It is going to be convenient to represent the most general
group element $g\in SL(2,\IR)$ in the interior of each of
the regions\foot{On the boundaries between the regions one
has to use a different representation (we shall return to this later).}
as
\eqn\grepr{g(\alpha,\beta,\theta;\epsilon_1,\epsilon_2,\delta)=
e^{\alpha\sigma_3}(-1)^{\epsilon_1}(i\sigma_2)^{\epsilon_2}
g_\delta(\theta)e^{\beta\sigma_3}~,}
where $\epsilon_1,\epsilon_2=0,1$; $\delta=I,1,1'$;
\eqn\gI{g_I=
\left(\matrix{
\cos\theta&-\sin\theta\cr
\sin\theta&\cos\theta\cr}\right);\qquad 0\le\theta\le{\pi\over2}~,}
\eqn\gone{g_1=g_{1'}^{-1}=
\left(\matrix{
\cosh\theta&\sinh\theta\cr
\sinh\theta&\cosh\theta\cr}\right);\qquad 0\le\theta<\infty~,}
and $\sigma_i$ are the Pauli matrices
\eqn\sss{\sigma_1=\left(\matrix{0&1\cr 1&0\cr}\right)~,\qquad
\sigma_2=\left(\matrix{0&-i\cr i&0\cr}\right)~,\qquad
\sigma_3=\left(\matrix{1&0\cr 0&-1\cr}\right)~.}
Equation \grepr\ describes the behavior of the group elements in all
twelve regions in figure 1. For example, $\epsilon_1=\epsilon_2=0$
corresponds to the regions $I,1,1'$. In $PSL(2,\IR)$ and the Poincare
patch, $\epsilon_1$ is taken to be identically zero.

Matrix elements of $g$ in a representation with
Casimir $-j(j+1)$, $K(j;g)$, are eigenfunctions
of the Laplacian with eigenvalue $-j(j+1)$.
We will discuss two types of unitary representations.
Principal continuous representations have
\eqn\jcont{j=-{1\over2}+is;\qquad s\in\IR~,}
and are further labeled by a phase $\exp(i\pi\epsilon)$,
where $\epsilon=0$ in $PSL(2,\IR)$, $\epsilon=0,1$ in
$SL(2,\IR)$, and $\epsilon\in [0,2)$ for the universal
cover. The phase $\exp(i\pi\epsilon)$ corresponds to
the representation of the center of the corresponding group.
The second class is principal discrete representations, characterized
by real $j$, with
\eqn\jdiscr{j\in \IZ+\epsilon/2~.}
We will choose a basis of eigenvectors of the non-compact $U(1)$,
$g=\exp(\alpha\sigma_3)$. For unitary representations,
the corresponding eigenvalue is $\exp(2im\alpha)$, with $m\in\IR$.
In a given representation, $m$ can take any real
value. Moreover, for the continuous representations, there
are two vectors with the same value of $m$, which we distinguish
by $\pm$.

\lref\misner{C.W. Misner, ``Relativity Theory and Astrophysics I:
Relativity and Cosmology,'' ed. J. Ehlers, 
Lectures in Applied Mathematics, volume 8, 160.}

For the continuous representations in the
above basis, the non-vanishing
matrix elements of $g$ \grepr\ are given by\foot{We will sometimes
use the label $g$ both for the $2\times 2$ matrices \defho,
as well as their representations.}
\eqn\matelem{\eqalign{
&K_{\pm\pm}(\lambda,\mu;j,\epsilon;g)\equiv
\langle j,\epsilon, m,\pm|g|j,\epsilon,\bar m,\pm\rangle=\cr
&e^{2i(m\alpha+\bar m\beta)}e^{i\pi\epsilon_1\epsilon}
\langle j,\epsilon, m,\pm|(i\sigma_2)^{\epsilon_2}
g_\delta(\theta)|j,\epsilon,\bar m,\pm\rangle~,\cr
}}
where
\eqn\lmmu{\lambda\equiv -im-j;\;\; \mu\equiv -i\bar m-j~.}
These matrix elements appear in \vilenkin\ (for the group $SL(2,\IR)$).
We will next give their values in each of the six regions
of the Poincare patch. There are two independent
functions; we begin in region 1, where we choose them to be
$K_{++}$ and $K_{--}$.~\foot{Actually,
$K_{++}$ vanishes in region 4  and $K_{--}$ vanishes
in region 2' (see below) where one should consider instead
$K_{-+}$ or $K_{+-}$.}
They are related via:
\eqn\kkk{\eqalign{
&K_{--}(\lambda,\mu;j,\epsilon;g_1)=
K_{++}(-i\bar m +j+1,-im+j+1;-(j+1),\epsilon;g_1)\cr
&={B(-i\bar m+j+1,i\bar m+j+1)\over B(-im+j+1,im+j+1)} 
K_{++}(-im+j+1,-i\bar m+j+1;-(j+1),\epsilon;g_1)~.}}
In region 1:
\eqn\kpp{K_{++}(\lambda,\mu;j,\epsilon;g_1)=
{1\over 2\pi i}B(\lambda,-\lambda-2j)
{(1-y)^{j+{\lambda+\mu\over 2}}\over (-y)^{{\lambda+\mu\over 2}} }
F(\lambda,\mu;-2j;{1\over y})~,}
\eqn\kmm{K_{--}(\lambda,\mu;j,\epsilon;g_1)=
{1\over 2\pi i}B(1-\mu,\mu+2j+1)
{(1-y)^{j+{\lambda+\mu\over 2}}\over (-y)^{2j+1+{\lambda+\mu\over 2}}}
F(\lambda+2j+1,\mu+2j+1;2j+2;{1\over y})~,}
\eqn\yyy{y\equiv -\sinh^2\theta~,}
where $g_1$ is given in eq. \gone, $\lambda,\mu$ are given in \lmmu,
$B(a,b)$ is the Euler Beta function
\eqn\bfff{B(a,b)={\Gamma(a)\Gamma(b)\over\Gamma(a+b)}~,}
and $F(a,b;c;x)$ is the hypergeometric function ${}_2F_1$.

In region $1'$ one finds
\eqn\kppp{K_{++}(\lambda,\mu;j,\epsilon;g_{1'})=
K_{--}(\lambda,\mu;j,\epsilon;g_1)~,}
\eqn\kmmm{K_{--}(\lambda,\mu;j,\epsilon;g_{1'})=
K_{++}(\lambda,\mu;j,\epsilon;g_1)~,}
where $g_{1'}=g_1^{-1}$ is given in eq. \gone, and
$K_{++}(g_1), K_{--}(g_1)$ are given in eqs. \kpp, \kmm, respectively.

In region $I$:
\eqn\kppi{K_{++}(\lambda,\mu;j,\epsilon;g_I)=
{1\over 2\pi i}B(\lambda,\mu+2j+1)
{(-x)^{j+{\lambda+\mu\over 2}}\over (1-x)^j}
F(\lambda,\mu;\lambda+\mu+2j+1;x)~,
}
\eqn\kmmi{K_{--}(\lambda,\mu;j,\epsilon;g_I)=
{1\over 2\pi i}B(1-\mu,-\lambda-2j)
{(1-x)^{j+1}\over(-x)^{j+{\lambda+\mu\over 2}}}
F(1-\lambda,1-\mu;1-\lambda-\mu-2j;x)~,
}
\eqn\xxx{x\equiv -{\rm ctg}^2\theta~,}
where $g_I$ is given in eq. \gI.

In region $II$ one has:
\eqn\kpppi{K_{++}(\lambda,\mu;j,\epsilon;g_{II})=
K_{--}(\lambda,\mu;j,\epsilon;g_I)~,}
\eqn\kmmmi{K_{--}(\lambda,\mu;j,\epsilon;g_{II})=
K_{++}(\lambda,\mu;j,\epsilon;g_I)~,}
where $g_{II}=(g_I)^{-1}$ ($g_I$ is given in eq. \gI), and
$K_{++}(g_I)$, $K_{--}(g_I)$ are given in eqs. \kppi, \kmmi, respectively.

In region $2'$:
\eqn\kpptwop{K_{++}(\lambda,\mu;j,\epsilon;g_{2'})=
K_{-+}(-\lambda-2j,\mu;j,\epsilon;g_1)~,}
\eqn\kmmtwop{K_{--}(\lambda,\mu;j,\epsilon;g_{2'})=
K_{+-}(-\lambda-2j,\mu;j,\epsilon;g_1)=0~,}
where $g_{2'}=i\sigma_2 g_1$ ($g_1$ is given in eq. \gone),
and
\eqn\kmp{\eqalign{K_{-+}(\lambda,\mu;j,\epsilon;g_1)&=
{1\over 2\pi i}(1-y)^{j+{\lambda+\mu\over 2}}\cr
&\times \Big[B(\lambda,1-\mu)(-y)^{\lambda-\mu\over 2}
F(\lambda,\lambda+2j+1;\lambda-\mu+1;y)\cr
&+(-)^\epsilon B(-\lambda-2j,\mu+2j+1)(-y)^{\mu-\lambda\over 2}
F(\mu,\mu+2j+1;\mu-\lambda+1;y)
\Big]~,}}
with $y$ given in eq. \yyy.
Here and below the formulae are valid for $PSL(2)$ where $\epsilon=0$
as well as $SL(2)$ where $\epsilon$ is either $0$ or $1$.

Finally, in region 4 one finds:
\eqn\kpptwop{K_{++}(\lambda,\mu;j,\epsilon;g_4)=
(-)^\epsilon K_{+-}(-\lambda-2j,\mu;j,\epsilon;g_1)=0~,}
\eqn\kmmtwop{K_{--}(\lambda,\mu;j,\epsilon;g_4)=
(-)^\epsilon K_{-+}(-\lambda-2j,\mu;j,\epsilon;g_1)~,}
where $g_4=-i\sigma_2 (g_1)^{-1}=-g_1i\sigma_2$ 
($g_1$ is given in eq. \gone),
and $K_{-+}(g_1)$ is given in eq. \kmp.

The behavior of the wavefunctions on the two dimensional surfaces
separating the various regions, \ie, $g\in SL(2,\IR)$ one of whose 
matrix elements is equal to zero,
requires a special treatment. Any $SL(2,\IR)$ matrix with a vanishing
entry can be written as
\eqn\gzeroel{g=e^{\phi\sigma_3}(-1)^{\epsilon_1}(i\sigma_2)^{\epsilon_2}
g_\gamma(i\sigma_2)^{\epsilon_3}~,}
where $\epsilon_{1,2,3}=0,1$;
\eqn\gbzero{g_\gamma=\left(\matrix{1&0\cr \gamma&1\cr}\right)~; 
\qquad \gamma>0~.}
The wavefunctions on all the ``lines'' in fig. 1 can thus be 
obtained from, say,
\eqn\kbdry{K_{++}(\lambda,\mu;j,\epsilon;g_\gamma)=
K_{--}(\lambda,\mu;j,\epsilon;g^{-1}_\gamma)=  
{1\over 2\pi i}B(\lambda,\mu - \lambda)\gamma^{\lambda - \mu}~.}
Note that as 
$\varepsilon\equiv \lambda-\mu=i(\bar m-m) \to 0$, 
$K\sim \gamma^\varepsilon/\varepsilon$.
This indicates a logarithmic 
divergence of the wavefunctions on the boundary $b=0$
(and similarly, on the other boundaries between the various regions the
wavefunctions are logarithmically divergent either when $m=\bar m$ or 
$m=-\bar m$).

\lref\tseytlin{
A.~A.~Tseytlin,
``Conformal sigma models corresponding to gauged
Wess-Zumino-Witten theories,''
Nucl.\ Phys.\ B {\bf 411}, 509 (1994)
[arXiv:hep-th/9302083].
}

\lref\GiveonTQ{
A.~Giveon and D.~Kutasov,
``Comments on double scaled little string theory,''
JHEP {\bf 0001}, 023 (2000)
[arXiv:hep-th/9911039].
}

\lref\dvv{
R.~Dijkgraaf, H.~Verlinde and E.~Verlinde,
``String propagation in a black hole geometry,''
Nucl.\ Phys.\ B {\bf 371}, 269 (1992).
}

A particularly interesting wavefunction is the combination \dvv:
\eqn\uuu{U(\lambda,\mu;j,\epsilon;g)\sim
K_{++}-{\sin(\pi\mu)\over\sin(\pi\lambda)}K_{--}~.}
It is infinitely blue shifted as $b\to 0$ (the boundary of region 1):
$U(b\to 0)\sim b^{i(m-\bar m)}$, and hence any normalizable 
wave packet constructed as a superposition of $U$'s  with
different values of $m$ and $\bar m$ vanishes at $b=0$.
Its asymptotic behavior in region 1 (as $\theta\to\infty$) is:
\eqn\asu{U(\chi,\omega;j;\theta\to\infty)\sim
e^{2i\chi\phi}e^{-\theta}\Big[
e^{2i(\omega t+s\theta)}+R(j;m,\bar m)e^{2i(\omega t-s\theta)}
\Big]~,}
where $j$ and $s$ are related by \jcont, 
\eqn\wltp{t=\alpha-\beta~, \qquad \phi=\alpha+\beta~, \qquad
\omega=\half (m -\bar m)~,\qquad \chi=\half (m+\bar m)~,}
and
\eqn\rrr{R(j;m,\bar m)={\Gamma(-2j-1)\Gamma(j+1+im)\Gamma(j+1-i\bar m)
\over\Gamma(2j+1)\Gamma(-j+im)\Gamma(-j-i\bar m) }
~.}
For $\omega,s>0$, the combination $U$ looks like an incoming plane
wave in region 1, scattered from the line $b=c=0$ (where regions
$1,1',I$ and $II$ intersect). The damping factor $ e^{-\theta}$ is
canceled by a corresponding factor in the $SL(2,\IR)$ measure.
$R(j;m,\bar m)$ is the ``reflection coefficient'' of a plane wave
coming in from the boundary\foot{The reference to ``incoming wave,''
``outgoing wave'' and  ``reflection'' is more appropriate
for discussing the NW coset, to which we turn in the next section.}.
It is also equal to the two point function
of $V_{j;im,i\bar m}$, a primary field of $SL(2)_L\times SL(2)_R$
with ``spin'' $j$ in the $SL(2)_k$ WZW model in the large $k$ limit
(see eq. (3.6) in \GiveonTQ). Note also that the reflection coefficient
is a phase when $m=\bar m$. This seems to be related to the fact that
for $m\not=\bar m$, the wavefunction \uuu\ is non-analytic all
along $b=0$; this behavior
allows part of the wave to leak from region $1$ to region
$II$ through the $b=c=0$ line. 
For $m=\bar m$, each of the wavefunctions $K_{++}$ and $K_{--}$
separately has a logarithmic singularity at $b=c=0$, but the combination
$U$ is regular, leading to complete reflection from the line
$b=c=0$.

\lref\mo{
J.~Maldacena and H.~Ooguri,
``Strings in AdS(3) and SL(2,R) WZW model. I,''
J.\ Math.\ Phys.\  {\bf 42}, 2929 (2001)
[arXiv:hep-th/0001053].
}

In string theory on $AdS_3$, principal continuous 
representations actually describe ``bad tachyons'' which
are absent in infrared stable vacua\foot{Representations
with $j=-\half+is$ are also relevant for the description
of certain ``long string'' states \mo, but we are not going to
discuss them here.}. ``Good tachyons'' and massive states
correspond to principal discrete representations,
with real $j$.
The corresponding normalizable wavefunctions are given by
particular linear combinations of the $K_{\pm\pm}$ above.
Explicitly, in region 1, the normalizable wavefunctions are
\eqn\kppdis{K_{++}(g_1)\qquad {\rm if} \qquad j<-{1\over 2}~,}
\eqn\kmmdis{K_{--}(g_1)\qquad {\rm if} \qquad j>-{1\over 2}~,}
where $K_{++}(g_1)$ and $K_{--}(g_1)$ are given in eqs.
\kpp\ and \kmm, respectively, with $j\in\IR$.
The continuation of these functions to the other regions is normalizable
as well iff $j$ and $\epsilon$ are correlated as in eq. \jdiscr.
Actually, the functions in eqs. \kppdis, \kmmdis\ are not independent,
due to the relation \kkk. This is in agreement with the familiar
fact that only one of the two independent solutions of the Laplace 
equation on $AdS_3$ -- the universal cover of $SL(2)$ -- is normalizable.
These normalizable wavefunctions decay exponentially towards 
the boundary of $AdS_3$. Their counterparts -- observables in string theory --
are exponentially supported at the boundary, a well known fact in 
string theory on AdS \agmoo.

\newsec{The coset}

\subsec{Geometry}

The string theory background we are interested in has the form
\eqn\nwback{\left[SL(2,\IR)\times SU(2)/U(1)\times U(1)\right]
\times \MM_6~,}
where $\MM_6$ is a compact manifold, say $\MM_6=T^6$. We will
mostly discuss the four-dimensional (coset) part of the geometry,
which can be described as follows. Let $(g,g')$ be a point in
the product manifold $SL(2,\IR)\times SU(2)$. We identify all
the points on the $U(1)\times U(1)$ orbit parametrized by $(\rho,\tau)$,
\eqn\thon{(g,g')\to \left(e^{\rho\sigma_3}ge^{\tau\sigma_3},
e^{i\tau\sigma_3}g'e^{i\rho\sigma_3}\right)~.}
This is a special case of \NappiKV, who discussed a one parameter
set of coset CFT's labeled by a parameter $\alpha$. Equation \thon\
corresponds to $\alpha=0$ in the notation of that paper.

Parametrizing the $SL(2,\IR)$ matrix $g$ as in \grepr, and
similarly expressing the $SU(2)$ matrix $g'$ in terms
of Euler angles
\eqn\euler{g'(\alpha',\beta',\theta')=e^{i\beta'\sigma_3} e^{i\theta'\sigma_2}
e^{i\alpha'\sigma_3}~,}
the gauge transformation \thon\ takes the form
\eqn\gatr{\eqalign{\alpha \to& \alpha +
\rho \cr \beta\to& \beta + \tau\cr
\alpha'\to&\alpha'+\rho\cr \beta'\to&\beta' +
\tau\cr \theta \to& \theta\cr
\theta' \to& \theta'\cr \epsilon_1 ,
\epsilon_2, \delta \to&
\epsilon_1 , \epsilon_2, \delta~.}}
The points of the coset manifold are the orbits of the 
$SL(2)\times SU(2)$
manifold under \gatr. To get a picture of the structure of the coset
manifold one can fix the gauge by setting the two  $SL(2)$ coordinates
$\alpha$ and $\beta$ to zero. This leaves four coordinates,
the three compact coordinates $\alpha', \beta'$ and $\theta'$
of $SU(2)$ and the non-compact $\theta$  together with the
discrete coordinates $\epsilon_1, \epsilon_2$
and $\delta$, surviving from $SL(2)$. From this point of view 
the coset manifold is a continuous family of $SU(2)$ manifolds,
topologically three-spheres, depending on the parameter $\theta$,
for each of the regions described in the previous section.
Algebraically, this gauge corresponds to viewing the NW universe
as a $\theta$-dependent $J_3\bar J_3$ deformation of the $SU(2)$
WZW model \KiritsisFD\ (for a review, see \GiveonFU).

This description breaks down for $\theta=0$ or
$\theta={\pi\over 2}$ if $\delta =I$, where different regions meet.
The  $SL(2)$ matrices corresponding to these values of $\theta$
are fixed by some $U(1)$ subgroup of \uoneact\ and cannot be
used for a complete gauge fixing. At these points in $SL(2)$,
part of the gauge fixing has to be imposed on the $SU(2)$ part,
so the three-spheres which sit above $\theta =0$ or
$\theta={\pi\over 2}$ for $\delta =I$, are twisted
by gauge identifications.

We will use an alternative gauge fixing, $\alpha'=\beta'=0$.
In this gauge, the coset is labeled by the $SL(2)$ parameters
$\alpha,\beta,\theta$ and the discrete labels $\epsilon_1 ,
\epsilon_2, \delta$, as well as the (compact) $SU(2)$ coordinate
$\theta'\in [0,{\pi\over 2}]$. Each $0< \theta'< {\pi \over 2}$,
for which \euler\ is a good parametrization, corresponds to
a copy of the full $SL(2)$ manifold.

It is important to note that the gauge condition
$\alpha'=\beta'=0$ does not fix the gauge completely.
In the parametrization \euler\  of $SU(2)$, $\alpha'+\beta'$
and $\alpha'-\beta'$ are only defined modulo $2\pi$.
Thus the transformations \gatr\ with $\rho+\tau=2\pi n_1,
\rho- \tau = 2\pi n_2$,
with $n_1,n_2 \in \IZ$, preserve the gauge conditions on $\alpha'$ and
$\beta'$: a residual $\IZ\times \IZ$ gauge symmetry has survived the
 gauge fixing. This implies that in the $SL(2)$ copy sitting above
each $\theta'$ in the interval $(0,{\pi \over 2})$, a further
identification has to be made under \gatr\ with $\rho+\tau=2\pi n_1,
\rho- \tau = 2\pi n_2$. In the parametrization \defho\ of
$SL(2)$, this identification reads (see \uoneact),
\eqn\slid{\eqalign{a\to&a e^{2\pi n_1}\cr b\to&b e^{2\pi n_2}\cr
c\to&c e^{-2\pi n_2}\cr d\to&d e^{-2\pi n_1}~.}}
Notice that the points on the 
line $a=d=0$ are fixed under the $\IZ$ subgroup
corresponding to $n_2=0$, while those on the line $b=c=0$
are preserved by the $\IZ$ subgroup
$n_1=0$. As a result, we expect an orbifold singularity of the
coset manifold at these surfaces.  Note that each point with
$b=0$ or with $c=0$ is identical by \slid\ to a point arbitrarily
close to $b=c=0$. Similarly for $a$ and $d$.

Taking into account the identification \slid, the physical space corresponds
to a fundamental domain resulting from the division of the $SL(2)$ manifold,
sitting above each $\theta' \in (0,{\pi \over 2})$, by \slid. Such a
fundamental domain in $SL(2)$, for group elements corresponding to matrices 
$g$ \defho\ with non-zero entries $a,b,c,d$, can be chosen as the region
\eqn\fund{\eqalign{1\ge |{b\over c}|>&e^{-4\pi}
\cr 1\ge |{a\over d}|>&e^{-4\pi}~.}}
To that one should add
two regions with $b=0$ and $|c|$ arbitrarily small
 together with two similar intervals with $c=0$ and
$|b|$ arbitrarily  small. Two additional pairs of
such regions should be added in the neighborhood of
the lines $a=0$ and $d=0$. These regions form the fundamental
domain for the set of $SL(2)$ matrices with one vanishing
element. The identification results in a non-Hausdorff manifold, 
see {\it e.g.} \hawel. 

\bigskip
{\vbox{{\epsfxsize=2.3in
        \nobreak
    \centerline{\epsfbox{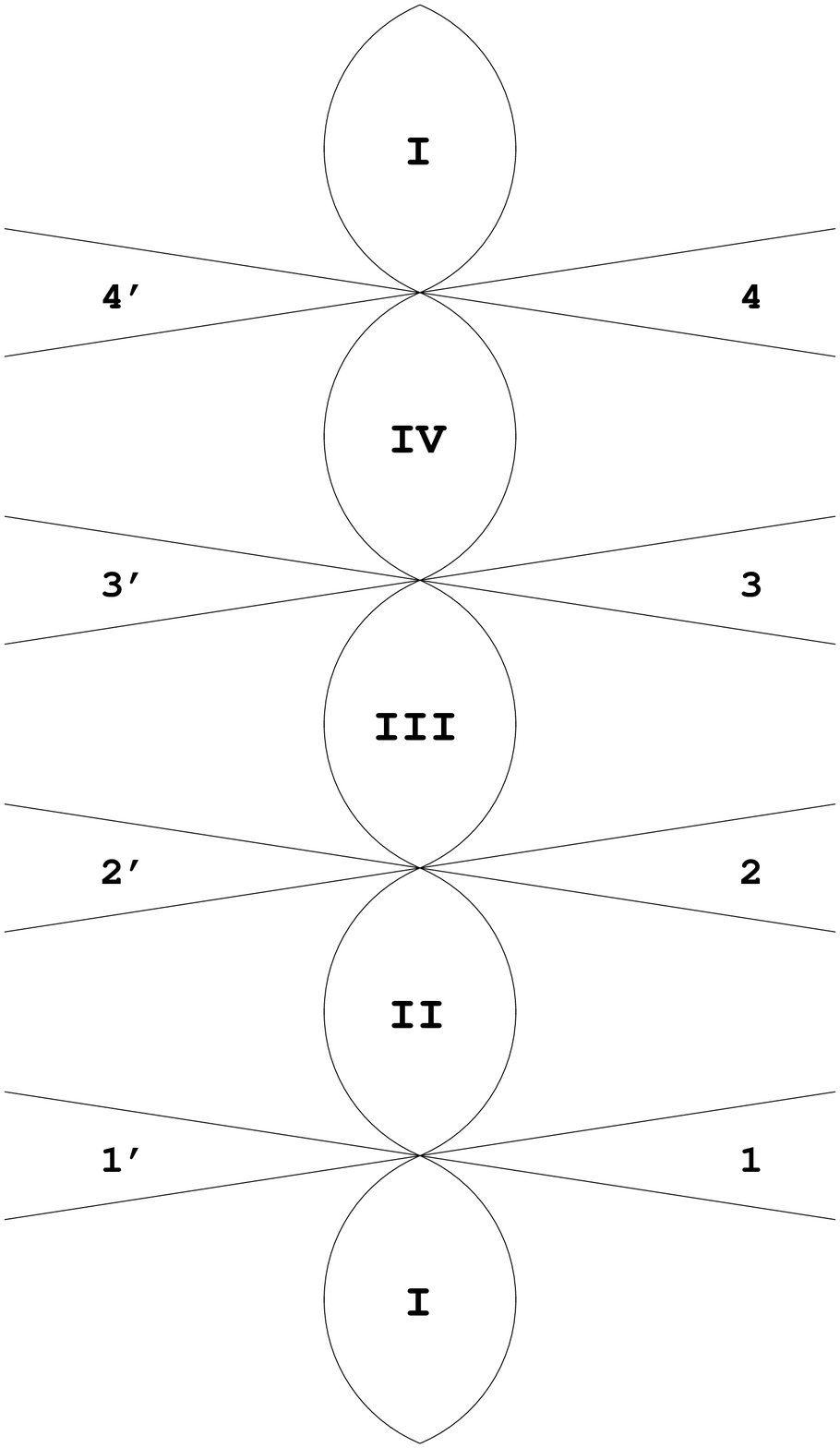}}
        \nobreak\bigskip
    {\raggedright\it \vbox{
{\bf Figure 3.}
{\it
A two dimensional slice of the four dimensional coset  spacetime.
In the ``closed universes'' -- regions $I,II,III,IV$ -- 
time ($\theta$) runs vertically. 
The horizontal axis represents $\lambda_\pm$, $\theta'$. 
In the ``whiskers'' -- regions
$1-4$, $1'-4'$ -- $\theta$ is spacelike and corresponds to the
horizontal axis. Time is either $\lambda_+$ or $\lambda_-$,
depending on the value of $\theta'$ (see figure 4). 
} }}}}
    \bigskip}
\noindent

The structure shown in figure 1 for $SL(2,\IR)$
turns after the identification \slid\ into that shown in figure 3.
Spacetime consists of a sequence of closed NW cosmologies connected
at the singularities, where they are also attached to additional regions
which were referred to above as whiskers. Near the big bang/big crunch
singularities, the manifold is non-Hausdorff. Near the line $b=c=0$,
the fundamental domain \fund\  has the form of four three-dimensional
wedges meeting at the line. It has there the form of a $(b,c)$ plane
divided by a finite boost (a Misner universe \misner), times a finite
interval in the coordinate $a$.~\foot{Misner background in string theory
appeared also in \KhouryBZ.} Similarly, near the line $a=d=0$ the modded
out $SL(2)$ manifold looks like the $(a,d)$ plane divided by a finite boost
times a finite interval in $b$. This three dimensional manifold sits above
each point $\theta' \in (0, {\pi \over 2})$, thus forming a four
dimensional universe.

To get the string frame metric, 
antisymmetric tensor and dilaton on this manifold,
one starts from the $SL(2)\times SU(2)$ sigma model and
introduces two $U(1)$ gauge fields $A^\rho$ and
$A^\tau$ corresponding to the two $U(1)$ identifications in \thon.
As usual \brs, the geometrical data of the coset manifold is
obtained by fixing  the gauge and integrating out these gauge fields.
We use the coordinates \grepr\ and \euler\ for $SL(2)\times SU(2)$
and fix the gauge $\alpha'=\beta'=0$.

In regions I, III, corresponding to $|W|<2$, with $W$ defined
in \defw, the procedure described above gives
\eqn\metrici{{1\over k}ds^2=-d\theta^2+d\theta'^2+{\cot^2\theta'\over 
{1+\tan^2\theta
\cot^2\theta'}}d\lambda_-^2 + {\tan^2\theta \over { 1+ \tan^2\theta \cot^2
\theta'}}d\lambda_+^2}
\eqn\bfi{B_{\lambda_+,\lambda_-}={k \over {1+\tan^2\theta \cot^2\theta'}}}
\eqn\dili{\Phi=\Phi_0-{1\over 2}\log(\cos^2\theta \sin^2 \theta' +
\sin^2\theta \cos^2\theta')}
where $\alpha \pm  \beta\equiv\lambda_{\pm}\in [0,2\pi)$,
and $\theta$ and $\theta'$ vary in the interval $[0,{\pi\over 2}]$.
In regions II, IV, $\lambda_+$ and $\lambda_-$ in \metrici\ switch their roles.
The parameter $k$ determines the maximal size of the universe (which is
$\sqrt{k}l_s$). In the CFT corresponding to the background \nwback, $k$ is the 
level of $SL(2)$ and $SU(2)$ (see section 4).~\foot{The 
geometric data obtained is valid in the large $k$ limit.
For the bosonic string there are known  $1/k$ corrections \BarsDX.
The exact background sometimes has a different singularity structure
\devega. For fermionic strings, the semiclassical background is
expected to be a solution to all orders in $1/k$ \refs{\BarsDX,\tseytlin}.} 
The dilaton $\Phi$ is normalized such that the string coupling is 
$g_s=e^{\Phi}$.

For the external regions corresponding to $W>2$ we find
\eqn\metrico{{1\over k}ds^2=d\theta^2+d\theta'^2+{\cot^2\theta'\over 
{1-\tanh^2\theta
\cot^2\theta'}}d\lambda_+^2 - {\tanh^2\theta \over { 1- \tanh^2\theta \cot^2
\theta'}}d\lambda_-^2}
\eqn\bfo{B_{\lambda_+,\lambda_-}={k \over {1-\tanh^2\theta \cot^2\theta'}}}
\eqn\dilo{\Phi=\Phi_0-{1\over 4}\log(\cosh^2\theta \sin^2 \theta' -
 \sinh^2\theta \cos^2\theta')^2}
where here $0 \le \theta < \infty ,\, 0\le \theta' \le {\pi \over 2} ,\,
\lambda_{\pm} \in [0,2\pi)$.
For $W<-2$, $\lambda_+$ and $\lambda_-$ in \metrico\ switch their roles.

The coordinates in \metrici, \bfi\ and \dili\ cover each of the
internal regions in fig. 3. Those of \metrico, \bfo\ and \dilo\ cover
an external region in the figure. In the internal regions, those
with $|W|<2$, \metrici\
implies that $\theta$ is a timelike coordinate varying over the finite
interval $[0,{\pi \over 2}]$. In the external regions $\theta$ becomes
spacelike; the timelike coordinate there is either $\lambda_-$ or
$\lambda_+$ depending on $\theta$ and $\theta'$ (see fig. 4). 
Since the coordinates $\lambda_{\pm}$ are periodic, the external regions
in fig. 3 -- the ``whiskers'' -- contain closed timelike curves.
Note also that the whiskers correspond to a time-independent background.
The scalar curvature in the whiskers is non-positive. On the other hand,
in the compact parts of spacetime the sign of the scalar curvature is position
dependent.

As mentioned in section 2, before gauging the $U(1)\times U(1)$, the
boundary of $AdS_3$ corresponds to large $|W|$ \defw; after gauging,
this corresponds to $\theta\to\infty$ in \metrico\ -- \dilo. 
In this limit the background factorizes. $\theta$ is described by
an asymptotically linear dilaton CFT; it is natural \AharonyUB\
to interpret $\theta\to\infty$ as a holographic screen. It is labeled
by $\{\theta',\lambda_\pm\}$, which form a three dimensional spacetime
with metric, $B$-field and dilaton given by:
\eqn\bndback{\eqalign{ {1\over k}
ds^2=&d\theta'^2+{\cot^2\theta'\over 1-\cot^2\theta'}d\lambda_+^2
-{1\over1-\cot^2\theta'}d\lambda_-^2\cr
B_{+-}=&{k\over1-\cot^2\theta'}\cr
\Phi=&\Phi_0-{1\over 4}\log(\cos2\theta')^2~.\cr
}}
This spacetime contains a timelike singularity at $\theta'=\pi/4$,
a kind of domain wall.

\bigskip
{\vbox{{\epsfxsize=2.3in
        \nobreak
    \centerline{\epsfbox{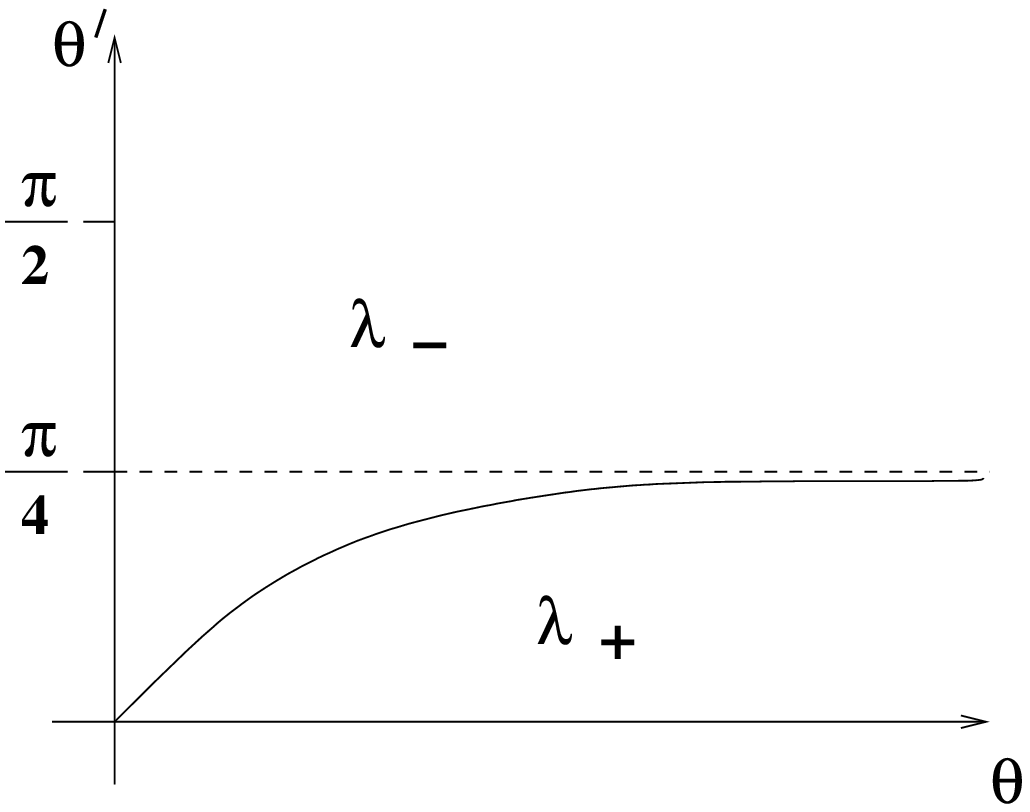}}
        \nobreak\bigskip
    {\raggedright\it \vbox{
{\bf Figure 4.}
{\it
The time coordinate in the metric \metrico\ valid
in regions $1-4$, $1'-4'$ depending on the value of
$\theta$ and $\theta'$. Note that for $\theta'>\pi/4$,
$\lambda_-$ is the time coordinate for all values of
$\theta$. For $\theta'<\pi/4$, $\lambda_-$ serves as
the time coordinate for $\cosh2\theta<1/\cos2\theta'$. 
} }}}}
    \bigskip}

\noindent
In an internal region, when the timelike coordinate $\theta$ tends to
$0$, the spatial metric in  \metrici\ shrinks to zero volume due
to the vanishing of the $d\lambda_+ ^2$ term. Similarly when  $\theta$
tends to ${\pi \over 2}$, the coefficient of $d\lambda_- ^2$ vanishes.
Such an internal region was interpreted
in \NappiKV\ as a four dimensional universe starting from a big bang at
time $\theta =0$ and ending at a big crunch at $\theta = {\pi \over 2}$.
Figure 3 describes then a series of closed universes; the big bang of each of
them is the big crunch of the previous one. Each such cosmological region
is connected at the singularity to two non-compact, time independent 
external regions with
closed timelike curves. In the universal cover of $SL(2,\IR)$, the geometry
contains an infinite number of copies of the structure exhibited in figure 3.

When $\theta=\theta'=0$, the metric \metrici\ develops a curvature
singularity and the dilaton \dili\ becomes singular. This corresponds
in $SL(2)\times SU(2)$ to
both $g$ and $g'$ in \thon\ being proportional to the identity matrix.
This is a fixed point of a $U(1)$ subgroup of \thon\ corresponding
to $\rho = -\tau$. Points preserved by a subgroup of the gauge group
give rise to singularities in the coset manifold. Similarly, \metrici\
and \dili\ are also singular at $\theta=\theta'={\pi \over 2}$. In
$SL(2)\times SU(2)$ this is the point corresponding to both $g$ and $g'$
proportional to the matrix $i\sigma_2$. This is fixed by the subgroup
of \thon\ with $\rho=\tau$.

\lref\GiveonSY{
A.~Giveon,
``Target space duality and stringy black holes,''
Mod.\ Phys.\ Lett.\ A {\bf 6}, 2843 (1991).
}

What attitude should one take towards the blowing up of the curvature
and coupling constant at some points? In general relativity this is a problem. 
On the other hand, a string theory based on a coset model is defined
algebraically by a perturbative genus expansion. The correlation functions
are inherited from those of the group manifold, which are regular.\foot{More
precisely, this is true for the unintegrated correlation functions, which are
the same in the CFT on the group manifold and in the coset CFT, up to
known functions of the worldsheet positions (see \eg\ \GiveonTQ). 
Thus, two and three
point functions are the same in the group manifold and in the coset CFT,
while for higher correlation functions, the integrated correlation
functions are in general different.} Hence strings can avoid the 
classical pathology, as might be expected for extended objects, even 
in a singular background \GiveonSY\ (for a review, see \GiveonFU).
A familiar example of this phenomenon is the geometric description of 
the $SU(2)/U(1)$ parafermion CFT as a bell-shaped manifold \bcr\
where both curvature and dilaton blow up at the edge. In that case
it is clear that string theory on $SU(2)/U(1)$ is perturbatively non-singular.

The background \metrici\ -- \dilo\ has two Abelian isometries.
As such, it is a particular point in the moduli space of more general
backgrounds related by the action of $O(2,2,\IR)$ rotations \GiveonKB\
(for a review, see \GiveonFU). A simple point in this moduli space
corresponds to the direct product of an $SU(2)/U(1)$ parafermion sigma
model with the Lorentzian $SL(2)/U(1)$ two dimensional black hole. Each
closed NW cosmological region denoted by a roman numeral in figure 3
is mapped by $O(2,2,\IR)$ to a region between the horizon and singularity
of the two dimensional black hole. The whiskers are mapped to the regions
outside the horizon and behind the singularity.

The discrete $O(2,2,\IZ)$ subgroup of $O(2,2,\IR)$ acts on this moduli space
as a T-duality group. In particular, there is an interesting T-duality
transformation which does not change the global shape of the NW universe but,
locally, removes the singularities at
$\theta=\theta'=0$ and $\theta=\theta'={\pi\over 2}$, creating instead
singularities at $\theta=0$, $\theta'={\pi\over 2}$ and
$\theta={\pi\over 2}$, $\theta'=0$.~\foot{This is similar to what
happens upon T-duality in the $SU(2)/U(1)$ sigma model; actually,
at the direct product point discussed above this T-duality is the 
one acting on the parafermion piece.} This is another reason
to believe that the physics of the NW model is non-singular at
$\theta=\theta'=0, {\pi\over2}$.

As mentioned earlier, the compact universe \metrici\ -- \dili\ can be 
described as a time-dependent $J\bar J$ deformation of an $SU(2)$ WZW 
model. In particular, the T-duality transformation which maps the 
deformation line of the $SU(2)_k$ WZW model to
$[{SU(2)_k\over U(1)}\times S^1_{\sqrt{k}R}]/\IZ_k$
(see \GiveonPH\ for details),
takes the cosmological CFT into an orbifold of the product of 
a parafermion with a time dependent circle, with radius 
$R(\theta)=\tan\theta$.
This equivalent CFT description might be useful
for exploring the theory further.\foot{In the vicinity of
$\theta=0$, the CFT background
$[{SU(2)_k\over U(1)}\times S^1_{\sqrt{k}R(\theta)}\times \IR_\theta ]/\IZ_k$
(where $\IR_\theta$ is timelike)
behaves like $[{SU(2)_k\over U(1)}\times {\IR^{1,1}\over \IZ}]/\IZ_k$.
This is the product of a two dimensional Misner universe
with a two dimensional parafermion sigma model (modded by 
a $\IZ_k$ which acts as a further boost
in $\IR^{1,1}/\IZ$ together with twisting the $\IZ_k$ parafermion CFT).}

For the external regions, there appears in \metrico\ -- \dilo\
a singular surface consisting of points with
\eqn\sing{\tan \theta' = \tanh \theta~.}
These do not correspond to any fixed points on the group
manifold. To understand the origin of this singularity recall that the
dynamics on the group manifold is governed by the quadratic form $E=G+B$
where $G$ is the metric and $B$ the antisymmetric tensor. The group manifold
is then modded out by identifying all the points on the orbit of the
 gauge group which is a two dimensional surface in the group manifold
parametrized by $\rho$ and $\tau$ of \thon. The singular points in \sing\
correspond to the orbits for which the quadratic form $E$ induced from
the full group manifold on the orbit becomes degenerate.
Indeed, the metric on the six dimensional $SL(2) \times SU(2)$ manifold,
in the region of $SL(2)$ with $W>2$,  is
\eqn\metrig{ds^2= d\theta^2 + \cosh^2 \theta
d\lambda_+^2 -\sinh^2\theta d \lambda_-^2 + d \theta'^2 + \cos^2 \theta'
d\lambda_+'^2 + \sin^2 \theta' d\lambda_-'^2~,}
where the coordinates \grepr\ and \euler\ are used with the substitution
$\lambda_{\pm}=\alpha \pm \beta$ and $\lambda_{\pm}'=\beta' \pm \alpha'$.
For $W<-2$,  $\lambda_+$ and $\lambda_-$ in \metrig\ switch roles.
The induced metric on a $(\rho,\tau)$ orbit is gotten from \metrig, using
\gatr, by substituting
\eqn\sorb{\eqalign{ d\lambda_+& = d\rho + d\tau\cr
d\lambda_-&= d\rho - d\tau\cr d\lambda_+'&= d\rho + d\tau \cr
d\lambda_-'&= d\tau - d\rho~.}}
The induced metric on the orbit is then
\eqn\metrorb{ds^2=2[d\rho^2 + d\tau^2 + 
(\cosh2\theta+\cos2\theta')d\rho d\tau]~.}
Notice that (unlike the regions with $|W|<2$) this gauged surface is 
not spacelike for large $\theta$.

The Wess-Zumino three-form for the
 $SL(2) \times SU(2)$ group is, for $|W|>2$,
\eqn\wz{{1 \over 3}Tr (g^{-1}dg)^3 ={1 \over 2}(\sinh2\theta \,d\lambda_+
 \wedge d\lambda_- \wedge d\theta - \sin 2\theta'\, d\lambda_+'
 \wedge d\lambda_-' \wedge d\theta')~.}
This gives for the $B$ field on the group manifold:
\eqn\bfg{ B=\cosh2\theta\,  d\lambda_+\wedge d\lambda_- + \cos2\theta'\,
 d\lambda_+'\wedge d\lambda_-'~.}
Substituting \sorb, the induced $B$ field on the gauge orbit is
\eqn\bforb{B=2(\cos2\theta'-\cosh2\theta)\,d\rho \wedge d\tau~.}
The induced quadratic form $E=G+B$ on the gauge orbit is the sum of
\metrorb\ and \bforb. In the $(\rho, \tau)$ coordinates it takes the form,
\eqn\eorb{E=2\left(\matrix{ 1 & \cos2\theta'\cr \cosh2\theta &1\cr}\right)~.}
This form degenerates when
\eqn\dete{\cosh 2\theta\, \cos 2\theta'=1~,}
which is the same as condition \sing\ which determines the singularities
in the external regions of the coset manifold.

\subsec{Wavefunctions}

Rather than studying the system as a sigma model describing string motion on
the complicated singular manifold presented in the previous subsection, we
will try to make use of its representation \NappiKV\ as a quotient of the 
much smoother $SL(2,\IR)\times SU(2)$ group manifold. The first step is to 
identify vertex operators on the group manifold, which are
invariant under the gauge identification. These give rise to vertex
operators in the quotient theory. A typical unexcited vertex operator on
$SL(2)\times SU(2)$ is of the form
\eqn\ver{V^{j,j'}_{m,m';\bar m, \bar m'}= K^j_{m,\bar m}(g)
D^{j'}_{m',\bar m'}(g')~,}
where  $K^j_{m,\bar m}(g)$ is a matrix element representing $g \in SL(2)$
in the unitary representation $j$ between vectors labeled by $m$ and $\bar m$.
$D^{j'}_{m',\bar m'}(g')$ is similarly defined for $SU(2)$.
For the gauging \thon,
it will be convenient  to choose for both the $SL(2)$ and $SU(2)$
representations a basis in which the $U(1)$ subgroup generated by $\sigma_3$
is diagonal. As in section 2, a vector labeled by $m$ is an eigenvector  of 
the operator representing the element $e^{\sigma_3 \rho}$ of $SL(2)$, with 
eigenvalue $e^{2im\rho}$. A vector labeled by $m'$ has the eigenvalue 
$e^{2im'\rho}$ for the operator representing the element $e^{i \sigma_3 \rho}$ 
of $SU(2)$. 
As described in the previous section, for $j=-\half+is$, corresponding 
to a continuous series representation of $SL(2)$, the representation
is not fully determined by $j$ but rather depends on an additional parameter
$\epsilon$. Also, in that case, the labels $m$ and $\bar m$ do not fully
specify a vector in the representation -- one needs to
further specify a $Z_2$ valued
index denoted above by $\pm$. Both  $K^j_{m,\bar m}(g)$ and
$V^{j,j'}_{m,m';\bar m, \bar m'}$ in eq. \ver\ depend 
on these additional parameters, which are omitted in \ver\ for brevity.

For  $g$ parametrized as in \grepr, $g(\alpha,\beta,\theta;\epsilon_1,
\epsilon_2,\delta)=
e^{\alpha\sigma_3}(-1)^{\epsilon_1}(i\sigma_2)^{\epsilon_2}
g_\delta(\theta)e^{\beta\sigma_3}$,  the dependence of the
matrix element $ K^j_{m,\bar m}(g)$ on $\alpha$ and $\beta$ is 
\eqn\kab{ K^j_{m,\bar m}(g)=
k^j _{m,\bar m,\epsilon_1 , \epsilon_2,\delta}(\theta)
e^{2i(m\alpha + \bar m \beta)}~.}
Similarly, for $g' \in SU(2)$ in the parametrization \euler,
$g'(\alpha',\beta',\theta')=e^{i\beta'\sigma_3}e^{i\theta'\sigma_2}
e^{i\alpha'\sigma_3}$, the element $D^{j'}_{m',\bar m'}(g')$ is of
the form
\eqn\dab{D^{j'}_{m',\bar m'}(g')=d^{j'}_{m',\bar m'}(\theta')
e^{2i(m'\beta'+\bar m' \alpha')}~.}
The operator $V^{j,j'}_{m,m';\bar m, \bar m'}$ is invariant under the
gauge transformation \gatr\ provided that
\eqn\gin{\eqalign{m=&\,-\bar m' \cr m'=&\, -\bar m}}
When this is satisfied, $V$ has the form
\eqn\vin{V^{j,j'}_{m,m';m',m} =  k^j _{m, m',
\epsilon_1 , \epsilon_2,\delta}(\theta)
d^{j'}_{m', m}(\theta')e^{im(\alpha - \alpha')}e^{im'(\beta-\beta')}~,}
which indeed depends only on the four gauge invariant coordinates $\theta,
\theta',\alpha-\alpha' ,\beta-\beta'$. 
These coordinates parametrize the coset
manifold. $V$ also depends on the discrete labels $\epsilon_1, \epsilon_2,
\delta$ which denote the different regions in fig. 3.

Although in unitary representations of $SL(2,\IR)$ the labels
$m$ and $\bar m$ take any real value, the coset condition \gin\ projects
out all $SL(2)$ operators except for those with integral or half
integral $m$ and $\bar m$, since the $SU(2)$ quantum numbers $m'$ and 
$\bar m'$ are obviously integral or half integral. Geometrically, the half 
integrality of $m$ and $\bar m$ guarantees that the vertex operator
$V$ remains single valued after the identifications \slid.~\foot{This 
may play an important role in ensuring that the model is consistent
even in the presence of regions with closed timelike curves.}
This is also valid along the surfaces on which a single element
of $g$ vanishes, where $K$ takes the form \kbdry.
This form embodies the special properties of that segment of space. 
In addition, $m$ and $\bar m$ are bounded by the requirement
\eqn\bomj{|m|,|\bar m | \le j'< {k\over2}~,}
where the latter condition comes from unitarity in the $SU(2)$ part
of \nwback. We will return to the spectrum of the theory in section 4,
where we also discuss excited string states;
here we note that \bomj\ implies a large depletion in the spectrum
of states that arises in going from $SL(2,\IR)\times SU(2)$ to the coset
\nwback. The spectrum of energies of single particle 
states\foot{Here we discuss the lowest lying string states; see section 4 
for the generalization to excited states.} is discrete and bounded from above 
(the energy in the whiskers is given by $m+\bar m$ or $m-\bar m$ depending on 
whether $\lambda_+$ or $\lambda_-$ is timelike; see fig. 4). 

Wavefunctions of the form \vin\ can be used to set up scattering
states in the whiskers part of the NW geometry and study their
dynamics. Consider, for example, an operator $V$ of the form
\eqn\vuu{V^{j,j'}_{m,m';m',m}=U(\lambda,\mu;j,\epsilon;g)
D^{j'}_{m',m}(g')~,}
where  $U(\lambda,\mu;j,\epsilon;g)$ is the combination of $SL(2)$ wave
functions discussed in eq. \uuu : $U(\lambda,\mu;j,\epsilon;g)\sim
K_{++}-{\sin(\pi\mu)\over\sin(\pi\lambda)}K_{--}~$.
Here\foot{Here and below we replace $m'\to -m'$ 
relative to the previous conventions, so that $\lambda$ and $\mu$ are 
treated symmetrically.} 
$\lambda = -im-j$ and $\mu =-i m'-j$, as in \lmmu;  $m$ 
and $m'$ are half integer
and $j$ is of the form $j= -{1 \over 2}+is$ with real $s$.
In region 1 of fig. 3, whose geometry is time independent,
$V$ describes a combination of an incoming and an outgoing
wave from the asymptotic region $\theta \to \infty$. As discussed 
in the previous section, on the group manifold $U$ was constructed 
in such a way that no flux enters region $1$ from region $I$.
This property is inherited by \vuu\ on the coset manifold.
The wave \vuu\ contains no flux coming from the adjacent 
cosmological region $I$ in fig. 3 through the big crunch of the
latter.

Without loss of generality, one can 
consider the case\foot{In other cases,
the assignments ``incoming/outgoing'' waves is different.}
when $\lambda_-$ is timelike, and sign$(s)=$ sign($\omega_-$), where
\eqn\omch{\omega_{\pm}=\half(m\pm m')~,}
so that $\omega_-$ is the energy, while $s$ and $w_+$ are components
of spatial momentum. In this case, $R(j;m,m')$ of \asu\ can be thought of 
as a reflection coefficient for a wave coming
from infinity in region 1, and scattering from the big crunch/bang at 
$\theta=0$. A part of this wave may penetrate  through the big bang 
into the cosmological region $II$. 
Indeed, 
\eqn\absr{|R|^2={\cosh(2\pi\omega_+)+\cosh(2\pi(s-\omega_-))\over
\cosh(2\pi\omega_+)+\cosh(2\pi(s+\omega_-))}~,}
which implies
\eqn\rleqone{|R|\leq 1~.}
In the coset, the transition from the cosmological regions to the 
whiskers and vice versa is possible only via the big
crunch/bang regions. The part of the incident wave that is not
reflected \absr\ apparently makes this transition from the whisker
to the cosmological region.

As in section 2, when $m=m'$, 
generically, the functions $K$ develop a logarithmic divergence
at the big crunch/bang of regions I/II, where whisker 1 is connected
(similarly, when $m=-m'$ they develop a singularity 
at a different big crunch/bang, say, the big crunch/bang of regions II/III). 
However, the combination $U$ is regular, and for the continuous
series representations $R(j;m,m)$ is a phase\foot{The
reflection coefficient also approaches a phase as $s$ tends to $0$.}. 
Therefore, for $m=m'$ the two independent wavefunctions in the continuous 
series can be chosen to be as follows.
One -- the combination $U$ --
is regular at the singularity, and describes a wave which is fully
reflected from it. The other
is logarithmically divergent at the singularity, and is not fully reflected.

A peculiar feature of the geometry of the 
whiskers is that the energy and momentum 
assignments for the incoming and outgoing 
waves depend on the value of the coordinate 
$\theta'$. According to eqs. \metrico, \dete\ and fig. 4, 
for  $\theta'>{\pi\over 4}$,
$\lambda_{-}$ serves as a time coordinate for all values of $\theta$; hence
the energy and momentum are identified as in \wltp. However, for
$\theta'<{\pi\over 4}$ the roles of $ \lambda_{-}$ and $ \lambda_{+}$
are reversed once $\cosh 2\theta \cos 2\theta'>1$. This leads to a 
discontinuity in the
assignment of momenta and energies. The sense of incoming and outgoing
waves does not change. 
As discussed in the previous subsection, the surface \dete\
at which the assignment of energy and momentum flips, 
corresponds to a timelike
singularity in the geometry \metrico\ -- \dilo. 
The wavefunctions are regular on this domain wall.
It is possible that this is due to the fact that 
this singularity does not originate from fixed points
of any gauge transformation 
(see the discussion at the end of subsection 3.1). 

So far we discussed the scattering process in 
region 1. Another interesting feature 
of the NW model is that, as we saw in section 2, 
the forms of the wavefunction
in the different regions in fig. 3 are not independent. 
In particular, specifying the
profile \vuu\ in region 1 produces some particular, 
non-trivial profiles of the field 
in other regions as well. 
This behavior can be determined by following the combination
$U$ given by eq. \uuu\ to the various regions, using eqs. \kppp\ -- \kmmtwop. 
For instance, in $PSL(2)$ ($\epsilon=0$), 
if the incoming wave in region 1 has weight 
one, the asymptotic behavior of $U(g_4)$, which follows from eq. 
\kmmtwop, implies that in region 4 there is an incoming wave with weight
$$-{\sin(\pi\mu)\over\sin(\pi\lambda)}
{\Gamma(-1-2j)\over B(\lambda,-\lambda-2j)}
\left[{\Gamma(1-\mu)\over\Gamma(-\mu-2j)}
+{\Gamma(\mu+2j+1)\over\Gamma(\mu)}
\right]~,$$
and an outgoing wave with weight
$$-{\sin(\pi\mu)\over\sin(\pi\lambda)}
{\Gamma(1+2j)\over B(\lambda,-\lambda-2j)}
\left[{\Gamma(-\lambda-2j)\over\Gamma(1-\lambda)}
+{\Gamma(\lambda)\over\Gamma(\lambda+2j+1)}
\right]~.$$
We see that specifying the boundary conditions near the
boundary of region 1 (at infinity) to correspond to the
wavefunction $U$ \uuu, actually describes a more complicated
process, with incoming and outgoing waves that are in general
non-zero in the different whiskers and cosmological regions
connecting them.\foot{In the covering groups, one may use 
combinations of waves corresponding to different $\epsilon$'s 
to eliminate the incoming waves in some of the whiskers.}
The total incoming flux from all regions equals the outgoing flux. 
This is guaranteed by current conservation on the group manifold. 
Since the flux  is conserved for every wave on the group,
it is also conserved for the restricted values allowed by the coset.
 
The correlation between the behaviors of
the wavefunctions in different whiskers
might be interesting for studying the
question of possible violations of causality
in this background due to the presence of
closed timelike curves, since one way to avoid
violations of causality is to impose constraints
on the Cauchy data in the theory. We see that such
constraints arise naturally in the NW model. One might
expect that correlations between the different
regions will lead to an effective non-locality in
the physics seen by any given observer. This is
an interesting issue that deserves further study.

Finally, following the discussion in section 2,
wavefunctions in the discrete series representations
decay exponentially towards the boundaries of the whiskers.
They are localized in the compact universes and their vicinity,
hence describing states ``living'' in the expanding and contracting 
cosmologies. The physics of these states is encoded in correlation
functions of the non-normalizable observables with real $j$ familiar
from AdS and LST. The principal discrete series states 
also interact with the scattering waves 
discussed above in the vicinity of the big bang/crunch, and in
the cosmological regions.
We will leave a more detailed analysis of these interactions,
as well as correlation functions of non-normalizable operators,
to future work.

\newsec{An algebraic analysis of the superstring on the NW background}

In this section we describe type II string quantization in the NW
background. We start, as a warmup, with a closely related
type II background,
\eqn\slsu{{SL(2,\IR)\over U(1)}\times {SU(2)\over U(1)}\times
T^6~.}
As discussed above, this background is in fact related to the
NW spacetime by an $O(2,2,\IR)$ transformation.
The first factor in \slsu\ is the two dimensional Minkowski
black hole; the second is an $N=2$ minimal model. The levels
of $SL(2)$ and $SU(2)$ in \slsu\ must be equal, and will be
denoted by $k$.

Consider first the $SL(2,\IR)$ factor in \slsu. $U(1)_L\times U(1)_R$
acts on the $SL(2,\IR)$ group element $g$ as
\eqn\galbt{g\to e^{\alpha\sigma_3}ge^{\beta\sigma_3}~,}
which is a spacelike non-compact direction in the group
manifold. In the usual conventions (see \eg\
\GiveonNS), it is generated by
the current $(iJ_3, i\bar J_3)$, where
$J_3=i\Tr \sigma_3g^{-1}\partial g$ satisfies
\eqn\normalz{J^3(z) J^3(0)=-{k\over2z^2}~.}
If one couples this current to a gauge field via
\eqn\lgauge{\CL_{\rm gauge}=i\bar A J_3+iA\bar J_3~,}
and integrates out the $SL(2,\IR)$ degrees of freedom,
one finds the effective Lagrangian for the gauge field
\eqn\Aact{\CL_A={k\pi\over2}\left(\bar A{\partial\over
\bar\partial}\bar A+A{\bar\partial\over\partial}A-2A\bar A
\right)~,}
which is invariant under the gauge transformation
$\delta A=\partial\alpha$, $\delta \bar A=\bar\partial\alpha$.
Imposing the gauge fixing condition $\partial_a A^a=0$, one
can parametrize the gauge field via a scalar field $t$,
\eqn\Adt{A=i\partial t;\;\;\bar A=-i\bar\partial t~.}
Plugging this into \Aact, we see that $t$ is dynamical,
and the full system consists of three parts:
\eqn\fullL{\CL=\CL_{SL(2)}+\CL_t+\CL_{\rm ghost}~,}
where (after rescaling $t$ to make it canonically
normalized)
\eqn\Lt{\CL_t=-\partial t\bar\partial t~.}
Note that $t$ is a timelike coordinate. Since the
gauge group \galbt\ is non-compact, $t$ is non-compact
as well.

$\CL_{\rm ghost}$ is the ghost Lagrangian
\eqn\Lghost{\CL_{\rm ghost}=b\bar\partial c+\bar b\partial\bar c~.}
The left-moving ghosts $b,c$ have scaling dimensions
$\Delta=1,0$, respectively, and similarly for the right movers
$\bar b$, $\bar c$.

The system \fullL\ has a BRST symmetry generated by
\eqn\qbrst{Q_{BRST}=\oint{dz\over 2\pi i}c\left(iJ_3+
i\sqrt{k\over2}\partial t\right)+
\oint{d\bar z\over 2\pi i}\bar c\left(i\bar J_3+
i\sqrt{k\over2}\bar\partial t\right)~.}
Physical states belong to the cohomology of $Q_{BRST}$.

So far we have discussed bosonic CFT on $SL(2)/U(1)$.
In the fermionic string there are also worldsheet fermions
in the adjoint representation of $SL(2)$, and the $U(1)$
that one gauges is a {\it super} affine Lie algebra generated
by the supercurrent
\eqn\superc{(\psi_3+\theta J_3,\bar\psi_3+\bar\theta \bar J_3)~,}
where $J_3$ is the total $U(1)$ current (it receives a contribution
from the fermions, $\psi^+\psi^-$), and $k$ is the level of the
full $SL(2)$ (which decomposes into bosonic and fermionic
contributions, via $k=(k+2)_B+(-2)_F$). There are also
bosonic ghosts $(\beta,\gamma)$, $(\bar\beta,\bar\gamma)$,
each with $\Delta=1/2$,
associated with gauging $i\psi_3$; $t$ has a superpartner
$\psi_t$.

The BRST charge \qbrst\ receives a contribution
\eqn\qprime{\oint {dz\over 2\pi i}\gamma\left(i\psi_3+
i\sqrt{k\over2}\psi_t\right)+
\oint {d\bar z\over 2\pi i}\bar\gamma\left(i\bar\psi_3+
i\sqrt{k\over2}\bar\psi_t\right)~.}
In order to describe the string theory on \slsu\ we have
to combine $SL(2,\IR)/U(1)$ with the other factors,
$SU(2)/U(1)$ and $T^6$. $SU(2)/U(1)$ is an $N=2$ minimal
model, but for our purposes it is convenient to describe
it in a way similar to $SL(2)/U(1)$ above. One starts with
$SU(2)$, and introduces a gauge field which couples to
the Cartan generator $K_3$, and gives rise, as in \Adt, to
a scalar superfield $(X,\psi_x)$, which is 
compact (like the gauge symmetry it is associated with).
There are also new ghosts, $(b', c')$ and $(\beta', \gamma')$,
which are analogs of \Lghost, \qbrst, \qprime.

To study type II string theory on \slsu, one also has
to apply a chiral GSO projection. While the background
\slsu\ has $N=2$ superconformal symmetry on the worldsheet,
it does not seem to be spacetime supersymmetric, since the
$U(1)_R$ charges are not integer (they are imaginary). 
A natural proposal for the action of GSO is:
$\psi\to -\psi$ for all fermions (those associated with
$SL(2,\IR)\times SU(2)\times T^6$, as well as $\psi_t$,
$\psi_x$) and $(\beta,\gamma,\beta',\gamma')\to-
(\beta,\gamma,\beta',\gamma')$ for the bosonic ghosts.
The question of spacetime SUSY in this language is the
question whether there exists a holomorphic, dimension
$(1,0)$ operator, $J_\alpha(z)$, which can be used to
form a spacetime supercharge
\eqn\qqaall{Q_\alpha=\oint{dz\over 2\pi i} J_\alpha(z)~.}
A natural conjecture here would be
\eqn\jalphadef{J_\alpha(z)=e^{-{\varphi\over2}}S_\alpha
e^{-{\varphi_1\over2}-{\varphi_2\over2}}~,}
where $\varphi$, $\bar\varphi$ are the bosonized superconformal
ghosts, $S_\alpha(z)$ is a spin field for the fourteen
worldsheet fermions associated with $SL(2)\times SU(2)\times T^6$,
and $\psi_x$, $\psi_t$, and $\varphi_1$, $\varphi_2$ are related
to the bosonic ghosts $\beta$, $\gamma$, $\beta'$, $\gamma'$ via
\eqn\bosonz{\beta\gamma=\partial\varphi_1;\;\;
\beta'\gamma'=\partial\varphi_2~.}
The total dimension of $J_\alpha$ is
\eqn\dimJa{\Delta(J_\alpha)={3\over8}+7\times{1\over8}-2\times{1\over8}=1~,}
as needed for spacetime SUSY, but while \jalphadef\ is invariant under
\qprime\ (this imposes some constraints on the spinor index $\alpha$),
it is not invariant under \qbrst\ (since all $J_\alpha$ are charged under
the CSA generator of $SL(2,\IR)$, $J_3$). 
Hence, the spacetime theory is not supersymmetric, 
which is not unexpected in a time-dependent background.
We have not proven that no other
supercurrents $J_\alpha(z)$ exist, but 
it is natural to expect that this is indeed
the case.

We next discuss some aspects of the resulting low energy
spectrum. Due to the chiral GSO projection, the lowest
lying states in the spectrum are ``gravitons.''
In the $(-1,-1)$ picture they have vertex operators
\eqn\vergrav{
e^{-\varphi-\bar\varphi}
V_{j;im,im}e^{i\sqrt{2\over k}mt}
V'_{j';m',m'}e^{\sqrt{2\over k}m'X}
e^{i\vec k\cdot\vec y}
\xi_{\mu\nu}\psi^\mu\bar\psi^\nu~,}
where the notation is as follows. As before,
$\varphi$, $\bar\varphi$ are the bosonized superconformal
ghosts. $V_{j;im,i\bar m}$ is a primary of $SL(2)_L\times SL(2)_R$
affine Lie algebra, with scaling dimension
$\Delta=\bar\Delta=-j(j+1)/k$. $V'_{j';m',\bar m'}$
is a primary of $SU(2)_L\times SU(2)_R$
affine Lie algebra, with scaling dimension
$\Delta=\bar\Delta=j'(j'+1)/k$.
The exponentials of $t$ and $X$ are the dressing by the gauge fields. 
Gauge invariance under \qbrst\ and its $SU(2)$ analog implies 
$m=\bar m$ and $m'=\bar m'$.
$\xi_{\mu\nu}$ is a polarization
tensor, with $\mu,\nu$ running over the six directions along the
$T^6$, and the ``coset directions,'' $\psi^\pm$ in $SL(2)$
and $\chi^\pm$ in $SU(2)$. For polarizations in the ``coset''
directions $\mu=\pm,\pm'$, the values of $im$ in
\vergrav\ are actually shifted by $\pm 1$, so that the eigenvalue
of the total $J^3$ on the state is imaginary. This can be seen
by imposing the standard transversality conditions on the vertex
operator \vergrav. Thus, for example, the $\psi^+\bar\psi^+$
term in \vergrav\ in fact looks like
\eqn\modver{
e^{-\varphi-\bar\varphi}
V_{j;im-1,im-1}e^{i\sqrt{2\over k}mt}
V'_{j';m',m'}e^{\sqrt{2\over k}m'X}
e^{i\vec k\cdot\vec y}\xi_{++}\psi^+\bar\psi^+~.}

\noindent
The mass-shell condition satisfied by \vergrav\ is:
\eqn\massshellc{-{j(j+1)\over k}-{m^2\over k}+
{j'(j'+1)\over k}-{m'^2\over k}=0~.}
The right-movers give rise to a similar equation.
One can think of $m$ as the energy of the resulting state.
Equation \massshellc\ has two kinds of solutions.
For small enough $m$, the solution for $j$ is real. In this
case, one of the solutions of \massshellc\ corresponds
to a non-normalizable wave function \vergrav, which
is exponentially supported at the ``boundary'' of $SL(2)/U(1)$,
the region far from the horizon of the black hole. The other 
solution is normalizable, and describes a principal 
discrete series state.

For energies $m$ larger than a critical value that depends on
$j'$, $m'$, the solution of \massshellc\ has the form
$j=-\half+is$, and the corresponding wave function
is delta function normalizable. It can be thought of as describing
a scattering state of a graviton (or dilaton, or NS $B$ field)
coming in from infinity and scattering from the black hole,
as discussed in \dvv\ and in the previous sections.

\lref\SeibergBJ{
N.~Seiberg and S.~H.~Shenker,
``A Note on background (in)dependence,''
Phys.\ Rev.\ D {\bf 45}, 4581 (1992)
[arXiv:hep-th/9201017].
}

\lref\GiveonPX{
A.~Giveon and D.~Kutasov,
``Little string theory in a double scaling limit,''
JHEP {\bf 9910}, 034 (1999)
[arXiv:hep-th/9909110].
}

The non-normalizable observables \vergrav\ with $j\in \IR$, 
give rise to non-fluctuating couplings, or superselection sectors 
in the worldsheet Lagrangian \SeibergBJ.
In analogy with the case of $SL(2,\IR) (=AdS_3$) and Little String
Theory, one is led to interpret them as off-shell observables in a 
dual theory, and the region at infinity (the analog of \bndback) 
as a holographic screen  (as in \AharonyUB).~\foot{The correlation 
functions of the observables \vergrav\ with $j$ real are complex. 
It would be interesting to understand the implications of this better.}

One can also discuss excited string states in the same way.
At oscillator level $N$, the mass-shell condition \massshellc\
is replaced by
\eqn\massshellN{-{j(j+1)\over k}-{m^2\over k}+
{j'(j'+1)\over k}-{m'^2\over k}+N=0~.}
The qualitative picture is the same: below a certain
critical energy $m_0$ which depends on $N$, one finds
non-normalizable observables with $j\in\IR$ and principal
discrete series states. Above that
critical value, one has scattering states with $j\in -\half +is$. 
Using standard techniques, one can compute scaterring
amplitudes of the states with $j\in -\half+is$ and
correlation functions of the sources with $j\in \IR$.

We now move on to a discussion of the NW background
\nwback, \thon.
The two $U(1)$ symmetries that one gauges are in this case
generated by
\eqn\nwsymm{(iJ_3,\bar K_3);\;\;(K_3, i\bar J_3)~.}
One can repeat the analysis of the two dimensional black hole
above for this theory. The BRST charge \qbrst\ is replaced by
\eqn\qbrstone{Q_{BRST}=\oint{dz\over 2\pi i}c\left(iJ_3+
i\sqrt{k\over2}\partial t\right)+
\oint{d\bar z\over 2\pi i}\bar c\left(\bar K_3+
i\sqrt{k\over2}\bar\partial t\right)~,}
and similarly, for the other $U(1)$ in \nwsymm\ one has
\eqn\qbrsttwo{Q_{BRST}=\oint{dz\over 2\pi i}c\left(K_3+
i\sqrt{k\over2}\partial t'\right)+
\oint{d\bar z\over 2\pi i}\bar c\left(i\bar J_3+
i\sqrt{k\over2}\bar\partial t'\right)~.}
The analog of the graviton \vergrav\ is now
\eqn\gravnw{e^{-\varphi-\bar\varphi}V_{j;im,im'}V'_{j';m',m}
e^{i\sqrt{2\over k}(mt+m't')}e^{i\vec k\cdot\vec y}\xi_{\mu\nu}
\psi_\mu\bar\psi_\nu~,}
where we have implemented the gauge conditions
$iJ_3+\bar K_3=0$, $K_3+i\bar J_3=0$, and the physical
state condition \massshellc\ is implied. $j$ runs over
the principal continuous series, $j=-\half+is$, $s\in\IR$, 
and over the principal discrete series, 
$j\in \IR$, $-\half<j<\half(k-1)$ for normalizable states
\GiveonPX.

An important difference between \gravnw\ and \vergrav\
is that in \gravnw\ $m,m'$ run over a finite set of values,
$m,m'\in \half \IZ$, $|m|, |m'|\le j'\le \half k-1$. 
Thus, unlike
\vergrav, which describes an infinite number of states, 
in \gravnw\ there is a finite number of physical states.
This is interesting from the point of view of holography. In
the black hole case, the fact that the energy $m$ in \vergrav\
is continuous, signals the fact that the dual theory is
$0+1$ dimensional (\ie, it is quantum mechanics). For NW, the finite
number of observables \gravnw\ seems to suggest that the dual
theory is zero dimensional (perhaps a finite collection of points
or a topological theory).

At higher oscillator levels $N$, one still has a finite number of
observables at each level, but that number is larger the higher
the level. Given $N$, the eigenvalue of the $SU(2)$ generator
$K_3$ is bounded
by $|K_3|\leq j'+N$. Hence, following the same reasoning as above,
the $SL(2)$ quantum number $m$ is bounded by
$|m|\leq j'+N\leq {k\over 2}-1+N$. 
Thus, for given $N$,
the number of states is finite.

\lref\GiveonUP{
A.~Giveon and D.~Kutasov,
``Notes on AdS(3),''
Nucl.\ Phys.\ B {\bf 621}, 303 (2002)
[arXiv:hep-th/0106004].
}

Correlators of the operators discussed above can be computed by using
standard perturbative worldsheet techniques. An important part of the
calculation comes from the operators 
$V_{j;im;im'}$ from the underlying $SL(2)$
theory. For instance, the reflection coefficient
\rrr\ is given by a two point function on the sphere 
(see eq. (3.6) in \GiveonTQ\ \foot{We 
choose $\nu(k)=1$ in eq. (3.6) of \GiveonTQ\
(see \GiveonUP\ for a discussion on the freedom to make such a choice).}):
\eqn\tpfgk{\eqalign{
R(j;m,m';k)&\equiv\langle V_{j;-im;-im'}V_{j;im;im'}\rangle\cr 
&={\Gamma(1-{2j+1\over k})\Gamma(-2j-1)\Gamma(j+1+im)\Gamma(j+1-im')
\over\Gamma(1+{2j+1\over k})\Gamma(2j+1)\Gamma(-j+im)\Gamma(-j-im') }~.}}
In the semiclassical limit, $k\to\infty$, this correlator coincides with 
the reflection coefficient $R(j;m,m')$ given in eq. \rrr;
\tpfgk\ is valid for all $k$. 
Note that, since in the continuous series $j=-\half+is$, the factor
$\Gamma(1-{2j+1\over k})/\Gamma(1+{2j+1\over k})$ is a phase. 
Hence, the $1/k$ corrections only add a $j$ dependent
phase to the semiclassical reflection coefficient, and in
particular they do not modify \absr. 

It is in principle possible to use the
algebraic description to study interactions.
In particular, one can use the results of
section 4 in \GiveonTQ\ to compute
the tree level three point functions. These seem to be regular for 
imaginary values of the spacelike Cartan eigenvalues.

\newsec{Summary}

The main purpose of this paper was to study in more detail the NW model
\NappiKV, which describes a closed cosmology starting with a big bang
and ending at a big crunch. 

Our attitude to this problem was that since the model can be described
as a coset CFT, string propagation in this spacetime should make sense, 
and thus it can be used to study various conceptual and quantitative 
issues that arise in backgrounds with cosmological (big bang and big
crunch) singularities. Examples include the nature of observables in such 
spacetimes, the question whether one 
should continue past such singularities to
pre big bang and post big crunch regimes, and if so how solutions of the 
wave equations are matched across the singularity. A better understanding
of these issues is necessary for studying the interactions between different 
regions separated by cosmological singularities, and of the question whether
the existence of singularities necessarily leads to large quantum effects
(\ie\ breakdown of string perturbation theory). 

\vskip .1in
\noindent
Our main results are the following:

String theory on the coset spacetime describes a sequence of big bang/big crunch 
universes attached to each other at the singularities in the way indicated in fig. 
3. Additional non-compact regions are attached to the sequence of closed universes, 
at the big bang and big crunch singularities as well. These non-compact regions, 
which we referred to as whiskers, are static and have a boundary at infinity, near 
which they look like spacelike linear dilaton solutions. 

The observables of the theory are defined by studying the behavior
of the fields near the boundary, as discussed in \AharonyUB\ in the context
of Little String Theory. As in LST, there are two kinds of observables. One 
corresponds to scattering states incident on the geometry from the boundary;
the other corresponds to non-normalizable wavefunctions supported near the
boundary. Both kinds of observables provide information about the physics
in the bulk of spacetime, and in particular on the closed cosmological 
region. The non-normalizable wavefunctions correspond to states localized
in the compact universes and their vicinity.

The scattering states, which correspond to vertex operators in the
principal continuous series of $SL(2,\IR)$, are partially reflected from
the cosmological singularity and partially transmitted into the cosmological
region. The reflection coefficient for this process is given in eq. \absr.
This provides an example of non-trivial interaction between regions separated
by cosmological singularities.

The non-normalizable observables seem to give rise to the analog 
of off-shell Green functions in AdS space and LST. These Green functions
encode the local dynamics in the bulk of spacetime. The two
point function of these observables is given by the standard $SL(2)$ result
(see \eg\ eq. (3.6) in \GiveonTQ). Higher point functions of both kinds of
observables can in principle be computed as well using results from CFT on
$SL(2,\IR)\times SU(2)$.  
 
The question of matching of solutions of wave equations across cosmological 
singularities is important for applications (see \eg\ 
\refs{\VenezianoPZ,\KhouryBZ}) and is in general ambiguous in the framework of 
QFT in curved spacetime. On the other hand, here the description of the space 
as a coset of an underlying $SL(2,\IR)$ manifold provides an organizing principle 
that allows one to continue the wavefunctions past the singularities. In section 
2 we described the wavefunctions in $SL(2,\IR)$, and in section 3 we explained how 
they give rise to uniquely determined wavefunctions in all the regions of the extended 
NW spacetime. 

We found that the spectrum of the theory is significantly depleted compared to
other related examples. We presented evidence that this depletion is related
to the appearance of closed timelike curves in some regions of the extended
NW spacetime (the whiskers). It would be interesting to obtain a better 
understanding of this relation, and in particular of the (lack of) violations of 
causality in this model.

Many questions deserve further study. We argued that three point functions
on the sphere should be finite, using the results of \GiveonTQ.
It would be interesting to establish this in more detail. It would
also be important to understand the structure of four and higher point
functions, and in particular the effects of the cosmological singularities
on these correlation functions. 

Since the model is not supersymmetric, it is of interest to 
consider loop corrections, and in particular the question of stability
of the classical spacetime against quantum corrections. Other issues
that require better understanding are the role of the timelike
singularity \dete\ in the whiskers, the effect of the closed
timelike curves in the whiskers, and the local physics in the
closed cosmological regions.

\bigskip
\noindent{\bf Acknowledgements:}
We thank O. Aharony, T. Banks, J.F. Barbon, M. Berkooz, D. Berman,
B. Craps, S. Demir, M. Dine, D. Friedan, R. Livne, E. Martinec, T. Piran, 
G. Rajesh, A. Schwimmer, N. Seiberg, S. Shenker, E. Sorkin and 
E. Witten for discussions. Special thanks go to D. Kazhdan for 
his help and to Y. Oz for very valuable discussions. 
D.K. and E.R. thank the Rutgers NHETC for its hospitality.
This work is supported in part by BSF -- American-Israel Bi-National 
Science Foundation, the Israel Academy of Sciences and Humanities --
Centers of Excellence Program, the German-Israel Bi-National Science 
Foundation, the European RTN network HPRN-CT-2000-00122 and DOE
grant DE-FG02-90ER40560.

\listrefs
\end